\documentclass[12pt,a4paper]{article}

\usepackage{amsmath,amssymb,xcolor,pstool,tikz,tensor,amsthm,mathtools,soul,braket,amsbsy}
\usepackage{wasysym}
            
\usepackage{amsmath}
\usepackage{mathtools}
\usepackage{amsfonts}
\usepackage{amssymb}
\usepackage{graphicx}
\usepackage{color}
\usepackage{booktabs}
\usepackage{inputenc}
\usepackage[T1]{fontenc}
\usepackage{mathrsfs}
\usepackage{enumerate}
\usepackage{xcolor,url}
\usepackage{cancel,dsfont} 
\usepackage[normalem]{ulem}

\usetikzlibrary{decorations.markings,positioning}
\usetikzlibrary{decorations.pathmorphing}
\usetikzlibrary{decorations.markings,positioning}

\usepackage{amsmath,amssymb,xcolor,pstool,tikz,tensor,amsthm,mathtools,soul,braket}
\usepackage[colorlinks = true,
            linkcolor = blue!70!black,
            urlcolor  = red!70!black,
            citecolor = green!55!black,
            anchorcolor = blue!70!black,bookmarks]{hyperref}

\usepackage[top=0.5in, bottom=0.6in, left=0.5in, right=0.5in]{geometry}

\def\di{\mathrm{d}}
\usepackage{wasysym}

\def\red{\textcolor{red}}

\def\C{\mathbb{C}}
\def\R{\mathbb{R}}

\def\kz{\boldsymbol{k}_0}

\newcommand\ee{\end{equation}}
\newcommand\be{\begin{equation}}

\newcommand{\eea}{\end{eqnarray}}
\newcommand{\bea}{\begin{eqnarray}}

\def\be{\begin{equation}}
\def\ee{\end{equation}}

\def\nn{\nonumber}

\newcommand{\dd}{\partial}

\usepackage{tikz}
\usepackage{esint}

\usepackage{graphicx}
\usepackage{amsmath}
\usepackage{amssymb}

\makeatletter

\let\DOTSI\relax
\newcommand*{\letteronint}[1]{%
  \DOTSI
  \mathop{%
    \mathpalette\@LetterOnInt{#1}%
  }%
  \mkern-\thinmuskip
  \int
}
\newcommand*{\@LetterOnInt}[2]{%
  \sbox0{$#1\int\m@th$}%
  \sbox2{$%
    \ifx#1\displaystyle
      \textstyle
    \else
      \scriptscriptstyle
    \fi
    #2%
  \m@th$}%
  \dimen@=.4\dimexpr\ht0+\dp0\relax
  \ifdim\dimexpr\ht2+\dp2\relax>\dimen@
    \sbox2{\resizebox*{!}{\dimen@}{\unhcopy2}}%
  \fi
  \dimen@=\wd0 %
  \ifdim\wd2>\dimen@
    \dimen@=\wd2 %
  \fi
  \rlap{\hbox to \dimen@{\hfil
    $#1\vcenter{\copy2}\m@th$%
  \hfil}}%
  \ifdim\dimen@>\wd0 %
    \kern.5\dimexpr\dimen@-\wd0\relax
  \fi
}

\newcommand*{\curvearrowbotright}{\mathpalette\rotmath@internal\curvearrowleft}
\newcommand*\rotmath@internal[2]{\rotatebox{180}{$\m@th#1#2$}}

\makeatother

\begin{document}

\begin{center}

{\Large \bf Positivity bounds on effective field theories\\[0.3cm]  with spontaneously broken Lorentz invariance}  \\[0.7cm]

\large{Paolo Creminelli$^{\,\rm a, \rm b}$, Oliver Janssen$^{\,\rm a, \rm b}$ and Leonardo Senatore$^{\,\rm c}$}
\\[0.5cm]

\small{
\textit{$^{\rm a}$
ICTP, International Centre for Theoretical Physics\\ Strada Costiera 11, 34151, Trieste, Italy}}
\vspace{.2cm}

\small{
\textit{$^{\rm b}$
IFPU, Institute for Fundamental Physics of the Universe\\ Via Beirut 2, 34014, Trieste, Italy }}
\vspace{.2cm}

\small{
\textit{$^{\rm c}$ Institut f\"ur Theoretische Physik, ETH Z\"urich, 8093 Z\"urich, Switzerland}}
\vspace{.2cm}

\vspace{.6cm}
\end{center}

\hrule \vspace{0.3cm}
{\small  \noindent \textbf{Abstract} \noindent \vspace{0.1cm}

\noindent We derive positivity bounds on EFT coefficients in theories where boosts are spontaneously broken. We employ the analytic properties of the retarded Green's function of conserved currents (or of the stress-energy tensor) and assume the theory becomes conformal in the UV. The method is general and applicable to both cosmology and condensed matter systems. As a concrete example, we look at the EFT of conformal superfluids which describes the universal low-energy dynamics of CFT's at large chemical potential and we derive inequalities on the coefficients of the operators, in three dimensions, at NLO and NNLO. 
\vspace{0.3cm}}
\hrule

\vspace{0.3cm}
\newpage

\tableofcontents

\section{Introduction}
The logic of effective field theories (EFT's) is to write down all possible operators in terms of the relevant low-energy degrees of freedom: the symmetries of the system dictate which operators are allowed, while the details of the ultraviolet (UV) theory are encoded in the coefficients of the operators. These coefficients however are not completely free. Besides the obvious constraints that guarantee a healthy EFT (e.g. there is a ground state), there are other bounds which are not easy to guess from the low-energy perspective. Under the mild assumptions that the UV completion is Lorentz-invariant, local and unitary, one can derive various inequalities that the low-energy coefficients must satisfy. The crucial information is the analytic structure of the $S$-matrix: Cauchy's theorem allows one to relate an integral along a low-energy contour, which can be reliably computed in the EFT in terms of its coefficients, to a high-energy one and this second integral can be proven to have a definite sign by the optical theorem. These ``positivity bounds'' and the implied logic that ``not everything goes'' was put forward in \cite{Adams:2006sv} and then generalized in many subsequent papers, trying to separate EFT's that can and cannot have a conventional UV completion (for a list of references see the recent \cite{Bellazzini:2020cot,Caron-Huot:2020cmc,Tolley:2020gtv, Arkani-Hamed:2020blm} and references therein).

It would be very interesting to extend this program to include theories in which Lorentz invariance is spontaneously broken, because of the existence of a preferred frame.  This setup is the one typical of cosmology and of condensed matter physics. There are various reasons why this extension would be useful, e.g. in cosmology one could reduce the allowed parameter space by knowing which theories of inflation or dark energy admit a standard UV completion. Moreover, in cosmology one is often tempted to consider EFT's which are ``unconventional'' and might in fact be impossible to incorporate in a healthy UV complete theory. For instance, Galileons \cite{Nicolis:2008in} are EFT's which arise when trying to modify gravity on large scales \cite{Luty:2003vm,deRham:2010kj}, or when trying to produce unconventional cosmological evolutions, like bounces, which violate the null energy condition \cite{Nicolis:2009qm}. Already the original paper \cite{Adams:2006sv} points out that  Galileons are an example of a theory that violates the inequalities one finds requiring a standard UV completion. However, it is not clear whether the same arguments can be used to rule out Galileons in a cosmological setting, where the background solution is time-evolving and thus Lorentz-breaking. Similar considerations apply to the Ghost Condensate~\cite{Arkani-Hamed:2003pdi}, which is also a theory that modifies gravity on large scales, violates the null energy condition~\cite{Creminelli:2006xe}, producing non-standard cosmologies~\cite{Creminelli:2007aq}, but at the cost of being in tension with the standard interpretation of black hole thermodynamics~\cite{Dubovsky:2006vk}. The application of these arguments to condensed matter systems, including QFT's at finite temperature or chemical potential, would put limits on what can be engineered in the laboratory and by Nature. Of course in this case the assumptions about the UV are completely robust, since one wants these systems to be embedded in the Standard Model. 

There are various reasons why it is not completely straightforward to extend the ``positivity'' arguments to theories in which Lorentz symmetry is non-linearly realized. One could imagine starting from a Lorentz-invariant theory, putting constraints on its Lagrangian following the logic above and then ``move'' to a Lorentz-breaking state. However, in order to know the EFT of perturbations around the new vacuum one should have constraints on the full non-linear Lagrangian in the Lorentz-invariant state. For instance one should study all operators $(\partial\phi)^n$ for arbitrary $n$, because they will all contribute to a $2 \to 2$ scattering once we consider a time-dependent solution. However the known positivity bounds have been derived mostly by considering $2 \to 2$ scattering and very little is known in the case of a scattering with a larger number of legs. This may simply be a technical limitation: more importantly, it is in general wrong to think about an EFT in which Lorentz symmetry is non-linearly realized as a Lorentz-invariant theory around another vacuum. In general the EFT makes sense only as a Lorentz-breaking theory and cannot be ``extrapolated'' to a Lorentz-invariant state. Consider for instance an ordinary fluid: the EFT that describes its waves is clearly not Lorentz-invariant and to turn off the Lorentz-breaking one should send to zero the density of the unperturbed fluid, but in this limit the theory of its perturbation of course loses its meaning. The same can be said about inflation: there is no guarantee that the EFT of perturbations of inflation can be extrapolated to a Lorentz-invariant state -- the EFT may break down as one switches off the time dependence of the background.

A related difficulty arises in applying the standard positivity arguments directly to the system which is not Lorentz-invariant. The $S$-matrix that describes the scattering of the EFT degrees of freedom (for instance the waves of a fluid) cannot be extrapolated to the UV, since the states are intrinsically low-energy and without boost invariance one cannot relate them to states in the UV theory. Therefore it is not clear how to connect the low-energy EFT calculation to the UV, which is a crucial step in the positivity logic. Reference \cite{Grall:2021xxm} (see also \cite{Baumann:2015nta}) extrapolated the $S$-matrix arguments from the Lorentz-invariant to the Lorentz-breaking case. However one is forced to make assumptions about the UV behavior which, in our opinion, are not justified in general.

In this paper we follow another route to derive constraints on theories in which boosts are broken. The inspiration is that analyticity arguments have been used for almost 100 years in the study of propagation of light in a material -- a setup which is clearly not Lorentz-invariant -- in the form of Kramers-Kronig relations. Our idea is to look at the correlation function $\braket{J^\mu(-k)J^\nu(k)}$ of a conserved current $J^\mu$ (or the stress-energy tensor $T^{\mu\nu}$), the same one which is responsible for the effect of a material on the propagation of light. We are going to assume that in the deep UV the theory reaches a conformal fixed point, so that the correlation function of $J^\mu$ is fully fixed by conformal invariance. This UV limit together with the controlled analytic properties of the two-point function of $J^\mu$ allow us to run an argument similar to the one of the $S$-matrix and derive positivity bounds on the coefficients of the EFT. The general ideas are illustrated in \S\ref{sec:setup}.

To illustrate the method, in this paper we focus on a particular example. We study the EFT that describes the low-energy excitations of a conformal field theory (CFT) at finite chemical potential $\mu$ for an internal $U(1)$ symmetry. For our purposes this is a neat example since the UV conformal invariance is broken only by the chemical potential, which is also the source of breaking of Lorentz invariance. In some sense this ``conformal superfluid'' is the most symmetric case of Lorentz-breaking where the whole conformal symmetry, and not only Lorentz symmetry, is non-linearly realized in the EFT. These theories have been the subject of intense study recently since they contain all the universal features of CFT's when considering operators with a large $U(1)$ charge. We will introduce these theories in \S\ref{superfluidsec} (postposing to Appendix \ref{EFToper} the systematic construction of all the relevant operators), calculating the correlators $\braket{J^\mu(-k)J^\nu(k)}$ in \S\ref{JJsec} and the positivity bounds that follow in \S\ref{JJpossec}. In sections \ref{TTsec} and \S\ref{TTpossec} we do the same for $\braket{T^{\mu \nu}(-k) T^{\rho \sigma}(k)}$, ending with a summary of all bounds in \S\ref{summarybounds}. In \S\ref{sec:loops} we explain why loop effects are absent for the coefficients we consider, and in Appendix \ref{app:conservation} we discuss the conservation properties of the two-point functions of $J^\mu$ and $T^{\mu \nu}$. A different argument that employs a contour which remains in the upper-half complex plane is discussed in Appendix \ref{k0sec}, and we point out in Appendix \ref{RKKsec} that our inequalities can also be derived by a relativistic generalization of the Kramers-Kronig relations. We will check the inequalities we find in two examples of tree-level UV completion (\S\ref{3DUVsec} and Appendix \ref{UV2scalars}). Conclusions and future directions are discussed in \S\ref{sec:conclusions}.

\section{\label{sec:setup}Setup and formalism}
We aim at deriving positivity bounds on coefficients of operators that appear in low-energy EFT's in which Lorentz invariance is spontaneously broken. We start by reviewing the well-known results of \cite{Adams:2006sv}. This will allow us to understand some of the essential ingredients that are needed to obtain analogous bounds when Lorentz invariance is spontaneously broken.

Ref.~\cite{Adams:2006sv} considers the $S$-matrix of $2\to 2$ scattering in the forward limit ({\it i.e.} $t\to 0$, where $s, t$ and $u$ are the Mandelstam variables). The $S$-matrix in this limit has some properties that allow one to derive positivity bounds (see~\cite{Bellazzini:2020cot} for a list of useful references):
\begin{enumerate}

\item\label{property1} It is a physically well-defined function for all real $s$. 

\item It is field redefinition independent.

\item It has an analytic continuation to the upper and lower half complex $s$-planes, with singularities residing only on the real axis, including unitarity cuts for energies $|s| > 4m^2$ where $m$ is the mass gap in the theory, which is assumed to be non-zero. This property is a consequence of locality and Lorentz invariance. 

\item The discontinuity across the cut on the positive real axis is $i\ \times$ a positive number. This is a consequence of unitarity.

\item It satisfies a crossing symmetry: {${\cal{M}}(s)^* = {\cal M}(4m^2-s^*)$}. This is a consequence of locality and Lorentz invariance.

\item It decays as $|\mathcal{M}(s)|/s^2 \rightarrow 0$ as $|s| \rightarrow \infty$. This property follows from the minimal requirements to derive the Froissart bound \cite{Jin:1964zza}.
\end{enumerate}
As in~\cite{Bellazzini:2020cot}, define $\hat s=s-2m^2$ and $\hat {\cal{M}}(\hat s)\equiv {\cal{M}}(s)$, so that $\hat{\mathcal{M}}(\hat{s})^* = \hat{\mathcal{M}}(-\hat{s}^*)$ by point 5 above. In the low-energy EFT, at tree level, the $S$-matrix takes the following polynomial form
\be
\hat{\cal M}(\hat{s}) = c_0 + c_2 \frac{\hat s^2}{\Lambda^4}+ c_4 \frac{\hat s^4}{\Lambda^8}+\ldots \ ,
\ee
where $\Lambda$ represents the scale suppressing the higher-dimension operators in the EFT, and $c_i$'s are real numbers.
 
\begin{figure}[h!]
\centering
\begin{tikzpicture}[thick,scale=0.8]

  \draw[->,gray] (-5,0) -- (5,0);
  \draw [->,gray] (0,-5) -- (0,5);

  \draw[gray, thick] (5,4) -- (5.5,4);
  \draw[gray, thick] (5,3.987) -- (5,4.5);
  \node at (5.26,4.32) {$\hat{s}$};

        \draw[black,
        decoration={markings, mark=at position 0.2 with {\arrow{>}}},
        postaction={decorate}] (4.5,0.22) arc
    [
        start angle=0,
        end angle=180,
        x radius=4.5cm,
        y radius =4.5cm
    ];
    
        \draw[black,
        postaction={decorate}] (4.5,-0.22) arc
    [
        start angle=0,
        end angle=-180,
        x radius=4.5cm,
        y radius =4.5cm
    ];
    
        \draw[black,
        postaction={decorate}] (-2.5,-0.22) arc
    [
        start angle=-90,
        end angle=90,
        x radius=0.22cm,
        y radius =0.22cm
    ];
    
        \draw[black,
        postaction={decorate}] (2.5,-0.22) arc
    [
        start angle=270,
        end angle=90,
        x radius=0.22cm,
        y radius =0.22cm
    ];
    
        \draw[black,
        decoration={markings, mark=at position 0.15 with {\arrow{>}}},
        postaction={decorate}] (0.35,0) arc
    [
        start angle=0,
        end angle=360,
        x radius=0.35cm,
        y radius =0.35cm
    ];

    \draw[black,decoration={
    markings,
    mark=at position 0.5 with {\arrow{>}}},
        postaction={decorate}] (-4.514,0.22) -- (-2.5,0.22);
    \draw[black,decoration={
    markings,
    mark=at position 0.5 with {\arrow{<}}},
        postaction={decorate}] (-4.514,-0.22) -- (-2.5,-0.22);
    \draw[black,decoration={
    markings,
    mark=at position 0.5 with {\arrow{>}}},
        postaction={decorate}] (2.5,0.22) -- (4.514,0.22);
    \draw[black,decoration={
    markings,
    mark=at position 0.5 with {\arrow{<}}},
        postaction={decorate}] (2.5,-0.22) -- (4.514,-0.22);
    
    \draw[decorate,decoration={zigzag, segment length=5, amplitude=2}] (2.5,0) -- (4.514,0);
    \draw[decorate,decoration={zigzag, segment length=5, amplitude=2}] (-2.5,0) -- (-4.514,0);
    
    \draw[black,fill=black] (0,0) circle (0.04cm);
    \draw[black,fill=black] (2.5,0) circle (0.04cm);
    \draw[black,fill=black] (-2.5,0) circle (0.04cm);

 \end{tikzpicture}
\caption{\label{fig:contour}The two contours in the $\hat s$-plane for $S$-matrix argument. The integral around the origin gives the coefficient of the operator in the EFT. It is equivalent to the large contour that reduces to an integral along the cuts, since the arcs at infinity vanish. The integral along the cuts is positive definite.}
\end{figure}
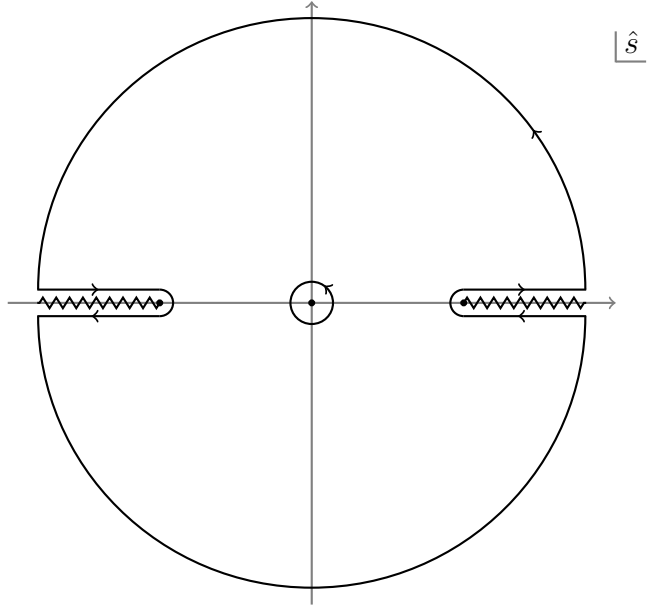

Now, following~\cite{Adams:2006sv}, we show how one can use these properties to derive positivity bounds. Consider the function $\hat{\cal{M}}(\hat{s})/\hat s^3$ and perform an integral counterclockwise around the origin. This function has a simple pole at the origin proportional to $c_2$ and so
\be
{\ointctrclockwise \di \hat s \;\frac{\hat {\cal{M}}(\hat{s})}{\hat s^3}=2\pi i \frac{c_2}{\Lambda^4}} \ .
\ee
By the analytic properties of $\hat{\cal{M}}/\hat s^3$, the contour can be deformed to another as in Fig.~\ref{fig:contour}. Because $\hat{\mathcal{M}}$ decays sufficiently quickly at infinity, the integral along the large circle is negligible as we send the radius to infinity. The integral along the negative-$s$ cut is equal to the one along the positive-$s$ cut, which gives $i \times c_+$, with $c_+$ a non-negative number. We therefore conclude that $c_2\geq 0$.

With this example in mind, let us now focus on theories where Lorentz invariance is spontaneously broken. The spontaneous breaking of Lorentz invariance (perhaps combined with the spontaneous breaking of other symmetries), leads to the presence of Goldstone bosons, which are the relevant degrees of freedom in these kind of EFT's. Typical examples are the phonons of solids, the sounds waves of fluids, or the Goldstone boson of time-translations in the effective field theory of inflation~\cite{Cheung:2007st} and of dark energy~\cite{Creminelli:2006xe,Creminelli:2008wc,Gubitosi:2012hu}. We will focus on cases where spacetime translations are effectively unbroken.\footnote{The low-energy unbroken spacetime translations are usually a linear combination of the original translations and some internal symmetry.}

Let us first point out an important obstacle in generalizing the $S$-matrix argument above to theories in which Lorentz symmetry is non-linearly realized. There is a crucial difference in the $S$-matrix depending on whether boosts are linearly realized or not. When Lorentz invariance is unbroken, the {\it in} and ${\it out}$ states are well-defined, no matter how large the value of $s$ is. In fact, the {\it in} and the {\it out} states are free single-particle states, and boost transformations relate low- and high-energy single-particle states. Therefore, one is guaranteed that the $S$-matrix is well-defined at all values of $s$: at low energy one can calculate it within the EFT, while at high energy the UV theory is needed.

When we deal with theories where Lorentz invariance is broken, this is no longer the case. As one increases the energy of the scattering, one is no longer guaranteed that the {\it in} states are well-defined, because there is no boost transformation that relates them to the low-energy states, which are the ones that are well-defined. For example one cannot arrange an experiment where two phonons scatter at, say, 2 TeV energy, because a single phonon state with 1 TeV energy simply does not exist. As the experimentalist increases the energy of the phonon more and more, at a certain point the description in terms of a single particle no longer applies and one probably has to deal with a complicated $n \to m$ scattering. Related to this, the EFT asymptotic states have a non-relativistic dispersion relation, while in the UV one expects to recover a relativistic dispersion. Therefore it is challenging to derive a connection between the low- and high-energy $S$-matrices. One could try to define an analytic function by analytic continuation of the low-energy EFT $S$-matrix, but one would not be guaranteed of any decay behaviour at infinity or positivity property.
The analytic properties of this function and its behavior at infinity could depend arbitrarily sensitively on the values at low energies.

Given these difficulties with the $S$-matrix, in this paper we explore another direction that we explain in the rest of this section.

\paragraph{IR/UV control.} As discussed above, we need a quantity whose behavior we can control both in the infrared (IR) and in the UV. Natural candidates are correlation functions of conserved currents\footnote{We will not consider accidental symmetries in the EFT.} (in particular we are going to stick to a $U(1)$ symmetry) or of the stress-energy tensor. This is what we will specialize on in this paper. We assume that, deep in the UV, our EFT's are completed by Lorentz-invariant, unitary CFT's.\footnote{This is a standard assumption if one wants a well-defined field theory, which does not require extra ingredients (like string theory) to be UV-completed~\cite{Komargodski:2011vj}.} For CFT's, conserved currents are primary operators with fixed scaling dimension. For a $d$-dimensional spacetime, the scaling dimension is $d-1$ for conserved currents and $d$ for the stress tensor. The two-point function of the currents is completely specified and in Fourier space it reads (see for instance \cite{Bzowski:2013sza})
\be\label{eq:JJCFT}
\braket{J^\mu(-k)J^\nu(k)} = c_J (k^\mu k^\nu - \eta^{\mu\nu} k^2) k^{d-4} \,,
\ee
where $c_J$ is a constant and we dropped the delta function of momentum conservation.\footnote{In $d = 4$ one gets $c_j \log k^2/k_0^2$. The presence of the scale $k_0$ indicates that there is a scale anomaly. Changing $k_0$ corresponds to changing the result by a polynomial in $k$, i.e.~a derivative of a delta function in real space. See \cite{Bzowski:2013sza}.} An analogous result holds for the stress-energy tensor. Notice that dealing with correlation functions of currents allows us to match another property of the $S$-matrix: correlation functions of Noether currents are field redefinition independent. We will assume the conserved currents are Hermitian operators.

Now that we have established that we can focus on correlation functions of currents, we need to choose which ones to study. Since we expect that locality and Lorentz invariance will play a crucial role (by inducing some form of analyticity that will allow us to perform integrations in the complex plane), a natural choice are the retarded and advanced Green's functions:
\begin{align}
	G_R^{\mu \nu}(x-y) &= i \theta(x^0-y^0) \braket{0|[J^\mu(x),J^\nu(y)]|0} \,, \label{GRdef} \\ \nn
	G_A^{\mu \nu}(x-y) &= -i \theta(y^0-x^0) \braket{0|[J^\mu(x),J^\nu(y)]|0} \,.
\end{align}

\paragraph{Analyticity.} Let us study the analytic properties of the retarded and advanced Green's functions. We define the Fourier transforms of the Green's functions as 
\begin{equation} \label{fourierG}
	\tilde{G}_{R,A}^{\mu \nu}(\omega,\boldsymbol{p}) = \int_{\mathbb{R}^d} \di^d x \, e^{-i p \cdot x} G_{R,A}^{\mu \nu}(x) \,.
\end{equation}
Our metric convention is $\{-,+,\ldots,+\}$. We have $G_R^{\mu \nu}(x) = 0$ for $x^0 < 0$, because of the $\theta$-function, and for $x^2 > 0$, since quantum fields commute at spacelike separation. So we may restrict the region of integration in~\eqref{fourierG} to the forward light cone (\text{FLC}): $x^0 > 0, x^2 < 0$. In the following, we will be considering complex values of the four-momentum $p$. Assuming polynomial boundedness of real-space correlation functions~\cite{Itzykson:1980rh}, the integral~\eqref{fourierG} converges for $\text{Re}(-i p \cdot x) < 0$ or $p^\textsf{Im} \cdot x < 0$ as $|x| \rightarrow \infty$ in the \text{FLC}. This requires $p^\textsf{Im} \in \text{FLC}$: for $p^\textsf{Im} \in \text{FLC}$,  $\tilde{G}_{R}^{\mu \nu}(\omega,\boldsymbol{p})$ is analytic. Analogously, $\tilde{G}_{A}^{\mu \nu}(\omega,\boldsymbol{p})$ is analytic for $p^\textsf{Im}$ in the backward light cone. We will explore these regions of analyticity in terms of the single complex variable $\omega$, setting
\begin{equation} \label{choicep1}
	\boldsymbol{p} = \boldsymbol{k}_0 + \omega \boldsymbol{\xi} \,,
\end{equation}
where ${\boldsymbol{k}_0, \boldsymbol{\xi} \in \mathbb{R}^{d-1}}$ are constants, with $|\boldsymbol{\xi}| \equiv \xi < 1$, with $\omega^\textsf{Im} > 0$ for $\tilde{G}_R$ and $\omega^\textsf{Im} < 0$ for $\tilde{G}_A$.\footnote{The choice in Eq.~\eqref{choicep1} is actually the most general. The three complex functions $\boldsymbol{p}(\omega)$ must be entire, otherwise we would introduce additional singularities not present in the function $\tilde G^{\mu\nu}$. Moreover these functions must be polynomially bounded: if this were not the case, plugging these functions in the CFT result, Eq.~\eqref{eq:JJCFT}, one would have a growth greater than any polynomial in some direction. This must be avoided since, as in the $S$-matrix case, we will need to neglect a contour at infinity after dividing the function by some power of $\omega$. An entire, polynomially bounded function is a polynomial, so the three functions $\boldsymbol{p}(\omega)$ must be polynomials. Consider now for each function the largest monomial in the polynomial, which dominates at large $|\omega|$: $a_n \omega^n$. Since $p^\textsf{Im}$ must be timelike, one needs $|{\rm Im} \; a_n \omega^n| < |{\rm Im}\;\omega|$ for any complex $\omega$ with large enough modulus. This is possible only for $n \leq 1$; $p_i(\omega) = a_0 + a_1 \omega$ for each $i \in \{0,1,2\}$. That $a_0 \in \mathbb{R}$ and $a_1 \in \mathbb{R}$ follows from the timelike constraint in the limit $\omega^\textsf{Im} \rightarrow 0$ with $\omega^\textsf{Re} \neq 0$, and with that $|a_1| < 1$ follows from the same constraint. So the most general choice is Eq.~\eqref{choicep1}.} We will assume that both functions $\tilde{G}_{R,A}^{\mu \nu}(\omega)$ may be defined on the real line $\omega \in \R$ by taking the appropriate limits $\omega^\textsf{Im} \rightarrow 0^\pm$.

Now for each $\boldsymbol{k}_0, \boldsymbol{\xi}$, we define the following function on the whole complex $\omega$-plane:
\begin{align}\label{eq:Gdef}
	\tilde{G}^{\mu \nu}(\omega) = \left\{
    \begin{array}{ll}
       \tilde{G}^{\mu \nu}_R(\omega,\boldsymbol{p}) & \mbox{if } \omega^\textsf{Im} \geq 0 \,, \\
        \tilde{G}^{\mu \nu}_A(\omega,\boldsymbol{p}) & \mbox{if } \omega^\textsf{Im} < 0 \,,
    \end{array}
\right.
\end{align}
with $\boldsymbol{p}(\omega)$ as in~\eqref{choicep1}. We can show that $\tilde{G}^{\mu \nu}(\omega)$ is analytic on $\C \setminus \{ (-\infty,-m) \cup (m,\infty) \}$, with $m$ a positive mass. In fact, consider an $\omega \in \R$ (or better, the limit $\omega \pm i \varepsilon, \varepsilon \rightarrow 0$). Inserting the unit operator ${\mathds{1} = \sum_n \ket{P_n} \bra{P_n}}$, with the sum running over all the eigenstates of the translations $\hat P$ (\footnote{Here and in the following we are using the unbroken low-energy spacetime translations, in general a linear combination of the original translations and some internal generator. For the $U$(1) case we are going to discuss in this paper, the current $J^\mu$ commutes with the internal generator.}), we find 
\begin{align}  \label{GRAdiff}
	&\lim_{\varepsilon \rightarrow 0} ~ \left(\tilde{G}^{\mu \nu}(\omega+i\varepsilon) - \tilde{G}^{\mu \nu}(\omega-i\varepsilon)\right) = i\int_{\mathbb{R}^d} \di^d x \, e^{-i p \cdot x} \braket{0|[J^\mu(x),J^\nu(0)]|0} \\ \nn
	&= i\int_{\mathbb{R}^d} \di^d x \, e^{-i p \cdot x} \braket{0|J^\mu(x) \left( \sum_n \ket{P_n} \bra{P_n} \right) J^\nu(0)|0} - \left( \mu \leftrightarrow \nu, x \leftrightarrow 0 \right) \\ \nn
	&= i\int_{\mathbb{R}^d} \di^d x \, e^{-i p \cdot x} \braket{0| e^{-i \hat{P} \cdot x} J^\mu(0) e^{i \hat{P} \cdot x} \left( \sum_n \ket{P_n} \bra{P_n} \right) J^\nu(0) |0} - \left( \mu \leftrightarrow \nu , x \leftrightarrow 0 \right) \\ \nn
	&= i(2\pi)^d \sum_n \left\{ \delta^{(d)}(p-P_n) \braket{0|J^\mu(0)|P_n} \braket{P_n|J^\nu(0)|0} - \delta^{(d)}(p+P_n) \braket{0|J^\nu(0)|P_n} \braket{P_n|J^\mu(0)|0} \right\} \,.\end{align}
Now for simplicity we shall assume there is a mass gap, so $P_n^0 > m > 0$ (we will relax this assumption momentarily). Since $p^0 = \omega$, if $|\omega| < m$ the arguments of the delta functions never vanish and $\tilde{G}^{\mu \nu}$ is continuous as we cross the imaginary axis here. For $\omega > m$, only the first term contributes, while for $\omega < -m$ only the second term contributes. We conclude that the function $\tilde{G}^{\mu \nu}(\omega)$ is analytic on the doubly-cut plane $\C \setminus \{ (-\infty,-m) \cup (m,\infty) \}$.\footnote{This can be shown using Morera's theorem.}

\paragraph{Positivity of cut contributions.} We will use a contour argument similar to the Lorentz-invariant case, see Fig.~\ref{fig:contour}.  We therefore need to understand if the contribution from the discontinuity across the cut has a definite sign.

Integrating $\tilde{G}^{\mu \nu}$ around the $(m,\infty)$ cut in the clockwise direction (i.e.~integrating~\eqref{GRAdiff} from $m$ to $+\infty$), with some arbitrary powers of $\omega$ inserted in the integration measure, we get a contribution only from the $\delta^{(d)}(p-P_n)$ terms. It will be useful to contract $\tilde G^{\mu\nu}$ with two copies of a constant real vector $V_\mu$.\footnote{More generally one can take the $V^\mu$ as polynomials in $\omega$, in which case one should contract with $V_\mu(\omega) V_\nu(\omega^*)^*$. One needs $V^\mu(\omega)$ entire so as not to change the analytic properties of $\tilde G^{\mu\nu}$ and polynomially bounded in order to be able to neglect the contribution from infinity: again, a polynomially bounded entire function is a polynomial. Since we obtain no new bounds in this paper by considering non-constant or complex polarization vectors, we restrict ourselves to constants $V^\mu \in \mathbb{R}$ here. In \S\ref{JJpossec} non-constant polynomials with real coefficients are briefly considered, and in Appendix \ref{k0sec} the more general case is treated.} We obtain
\begin{equation} \label{GVVint}
	\frac{1}{(2\pi)^d} \underset{(m,\infty) \text{ cut}}{\int} \frac{\di \omega}{\omega^\ell} \, \tilde{G}^{\mu \nu}(\omega) V_\mu V_\nu = i \int_m^\infty \frac{\di \omega}{\omega^\ell} \sum_n \delta^{(d)}(p-P_n) ~ \left|\braket{P_n|J^\mu(0) V_\mu |0} \right|^2 \,,
\end{equation}
where we have used that the $V_\mu$ are real and the $J^\mu$ are Hermitian.\footnote{Obviously we are assuming that the theory is unitary: when discussing Euclidean theories, we are assuming reflection positivity, see for instance \cite{Rychkov:2016iqz}.} This is of the form $i \times (\text{positive})$. Integrating around the $(-\infty,-m)$ cut also in the clockwise direction, we get 
\begin{equation}\label{eq:cut:temp}
	\frac{1}{(2\pi)^d} \underset{(-\infty,-m) \text{ cut}}{\int} \frac{\di \omega}{\omega^\ell} \, \tilde{G}^{\mu \nu}(\omega) V_\mu V_\nu = -i\int_{-\infty}^{-m} \frac{\di \omega}{\omega^\ell} \sum_n \delta^{(d)}(p+P_n) ~ \left|\braket{P_n|J^\mu(0) V_\mu |0} \right|^2 \,,
\end{equation}
which is $i \times (\text{positive})$ for odd $\ell$ and $i \times (\text{negative})$ for even $\ell$. We will be interested in odd $\ell$'s where we know that the sum of contributions from both cuts is of the form $i \times (\text{positive})$.

\paragraph{Crossing symmetry and reality properties.} The functions $\tilde{G}^{\mu\nu}_{R,A}(p)$ satisfy a crossing symmetry property. For $p^\textsf{Im} \in \text{FLC}$ we have
\begin{align}
	\tilde{G}^{\nu \mu}_A(-p) &= -i \int_{\mathbb{R}^d} \di^d x \, e^{i p \cdot x} \theta(-x^0) \braket{0|[J^\nu(x),J^\mu(0)]|0} \notag \\
	&= -i \int_{\mathbb{R}^d} \di^d x \, e^{-i p \cdot x} \theta(x^0) \braket{0|[J^\nu(-x),J^\mu(0)]|0} \notag \\
	&= -i \int_{\mathbb{R}^d} \di^d x \, e^{-i p \cdot x} \theta(x^0) \braket{0|[J^\nu(0),J^\mu(x)]|0} = \tilde{G}^{\mu \nu}_R(p) \,, \label{eq:crossing}
\end{align}
where we have changed the variable of integration from $x\to -x$ and used translation invariance. In the particular case $\boldsymbol{k}_0 = \boldsymbol{0}$ in Eq.~\eqref{choicep1} this implies, for all $\omega \in \mathbb{C}$,
\be
\tilde G^{\mu\nu}(\omega) = \tilde G^{\nu\mu}(-\omega) \quad {\rm when}~\, \boldsymbol{k}_0 = \boldsymbol{0} \,.
\ee
The functions $\tilde{G}^{\mu\nu}_{R,A}(p)$ also have certain reality properties. Since $G^{\mu \nu}_R(x)$ is real for real $x$ (from~\eqref{GRdef}, use the Hermiticity of the $J^\mu$), its Fourier transform satisfies 
\begin{equation}
	\tilde{G}^{\mu \nu}_R(p) = \tilde{G}_R^{\mu \nu}(-p^*)^*
\end{equation}
where $p^\textsf{Im} \in \text{FLC}$. Combining this with~\eqref{eq:crossing} we obtain
\begin{equation}\label{starstar}
	\tilde{G}^{\mu \nu}_R(p) = \tilde{G}^{\nu \mu}_A(p^*)^* \,.
\end{equation}

\paragraph{EFT contact terms and gauging the symmetry.} So far we have discussed the properties of $\tilde{G}^{\mu\nu}$ in the full UV theory. Since our purpose is to put constraints on the coefficients of the operators of the low-energy EFT, we have now to discuss the calculation of $\tilde{G}^{\mu\nu}$ in the EFT. One can think of the EFT as the theory after the heavy degrees of freedom have been integrated out. In the procedure of integrating out heavy modes one generates contact terms: for example, a heavy propagator $\propto (p^2 + m^2)^{-1}$ expanded at low energy gives a polynomial in $p^2$. This polynomial is the Fourier transform of a sum of contact terms, i.e.~derivatives of a delta function. It is crucial to keep track of all these contact terms, since they are part of the low-energy expansion of $\tilde{G}^{\mu\nu}$, the function with the nice analytic properties and UV behavior we discussed above.

 It is often useful to calculate correlation functions of a given operator by coupling it to an external source. In the case of a conserved current this is a non-dynamical gauge field $A_\mu$ (and a non-dynamical metric $g_{\mu\nu}$ in the case of the stress-energy tensor). Notice that this $U(1)$ symmetry implies that one has a gauge symmetry for $A_\mu$ (when the $U(1)$ is spontaneously broken, the case of interest below, this gauge symmetry is spontaneously broken). Instead of computing correlation functions of $J^\mu$, one computes functional derivatives in the path integral of the gauged theory.\footnote{This is also the way to deal with anomalous contact terms in the CFT, see \cite{Bzowski:2013sza}.}

Let us be more explicit. First, let us write the retarded Green's function in terms of the time-ordered product of operators:
\be\label{eq:green_time}
G_R^{\mu \nu}(x-y) = i \theta(x^0-y^0) \braket{0|[ J^\mu(x),J^\nu(y)]|0}=i \braket{0|{\rm T}\{J^\mu(x)J^\nu(y)\}|0}-i \braket{0|J^\nu(y)J^\mu(x)|0}\ .
\ee
It is easy to argue that the contact terms generated by integrating out the heavy modes will only appear in the time-ordered product and not in the last term of the equation above. Indeed one can do exactly as in Eq.~\eqref{GRAdiff} to obtain
\be\label{eq:KLunordered}
 i\int_{\mathbb{R}^d} \di^d x \, e^{-i p \cdot x} \braket{0| J^\nu(0) J^\mu(x) |0} = i(2\pi)^d \sum_n \delta^{(d)}(p + P_n) \braket{0|J^\nu(0)|P_n} \braket{P_n|J^\mu(0)|0}  \,.
\ee
This shows that the value of the second term on the right-hand-side (RHS) in~\eqref{eq:green_time} at low energies and momenta is only affected by the low-energy states, i.e.~the states in the EFT; the heavy states do not contribute. On the contrary, the first term on the RHS of~\eqref{eq:green_time} involves a convolution in Fourier space, because of the product with the $\theta$-function in real space. Therefore also heavy modes contribute to the time-ordered correlation function at low energy.

It is useful to see this in terms of path integrals. The first term is simple since it is time-ordered:
\be \label{TPIusual}
\braket{0|{\rm T}\{J^\mu(x)J^\nu(y)\}|0}= \frac{1}{Z} \int {\cal{D}}\phi\; e^{i\,\int_{\mathbb{R}^d} \di^d x\; {\cal{L}}(\phi)}J^\mu(x)J^\nu(y) \,,
\ee
where $Z$ is the normalization
\begin{equation}
	Z = \int {\cal{D}}\phi\; e^{i\,\int_{\mathbb{R}^d} \di^d x\; {\cal{L}}(\phi)} \,.
\end{equation}
Here $\phi$ denotes the whole set of dynamical fields in the theory, both heavy and light.  (For ease of notation, we have   assumed that we can write the path integral in terms of the Lagrangian, as nothing would change if we were to work with the Hamiltonian.) The second term in~\eqref{eq:green_time} is a correlation function without any ordering prescription. It is less common to express this term using path integrals, which in their usual form give a time ordering of the inserted operators as in~\eqref{TPIusual}. However, correlation functions which are not ordered, or with all possible orderings, are relevant when one is interested in observables different from the flat space $S$-matrix (in cosmology for instance); see \cite{Haehl:2017qfl} for a comprehensive treatment.  Using the time-evolution operator $U(t,t_0)$ from $t_0 \rightarrow t$ one can write the correlator in terms of Schr\"odinger-picture operators
\be
\braket{0|J^\nu(y)J^\mu(x)|0}=\braket{0|U(+\infty, y^0) J_{(s)}^\nu(\boldsymbol{y})U(y^0,x^0)J_{(s)}^\mu(\boldsymbol{x})U(x^0,-\infty)|0} \ . 
\ee
Inserting twice the identity at fixed time
\begin{equation}
	\mathds{1} = \int {\cal{D}} \phi(\tilde{\boldsymbol{x}}) \, \ket{\phi(\tilde{\boldsymbol{x}})} \bra{\phi(\tilde{\boldsymbol{x}})} \,,
\end{equation} 
at $t =x^0$ and $t = y^0$, and then expressing each of the time-evolution operators by a path integral with appropriate boundary conditions,
\be
\braket{\phi(y^0,\tilde{\boldsymbol{y}})|U(y^0,x^0)|\phi(x^0,\tilde{\boldsymbol{x}})}= \int_{\phi(\tilde{\boldsymbol{x}})}^{\phi(\tilde{\boldsymbol{y}})} {\cal{D}}\phi\; e^{i\,\int_{x^0}^{y^0} \di^d x\; {\cal{L}}(\phi)}\ ,
\ee
we can write
\begin{align}
	\braket{0|J^\nu(y)J^\mu(x)|0} &= \frac{1}{Z} \int {\cal{D}} \phi(\tilde{\boldsymbol{x}}) \int {\cal{D}} \phi(\tilde{\boldsymbol{y}})\  J^\nu(\phi(y^0,\boldsymbol{y}))  J^\mu(\phi(x^0,\boldsymbol{x}))  \int_{\phi(\tilde{\boldsymbol{y}})}^{\rm}{\cal{D}}\phi_3 \;e^{ i\,\int_{y^0}^{+\infty} \di^d x\; {\cal{L}}(\phi_3)} ~\times \notag \\ 
& \int_{\phi(\tilde{\boldsymbol{x}})}^{\phi(\tilde{\boldsymbol{y}})}{\cal{D}}\phi_2 \; e^{i\,\int_{x^0}^{y^0} \di^d x\; {\cal{L}}(\phi_2)}\  \int_{\rm}^{\phi(\tilde{\boldsymbol{x}})}{\cal{D}}\phi_1 \;e^{ i\,\int_{-\infty}^{x^0} \di^d x\; {\cal{L}}(\phi_1)}\ .
\end{align}
Notice that $\phi(x^0,\boldsymbol{x})$ is fixed in terms of $\phi(\tilde{\boldsymbol{x}})$: $\phi(x^0,\boldsymbol{x}) = \left.\phi(\tilde{\boldsymbol{x}})\right|_{\tilde{\boldsymbol{x}}=\boldsymbol{x}}$ (similarly for $y^0$). The selection of the interacting vacuum works in the same way as in the standard T-ordered case, with an evolution in Euclidean time in the asymptotic past and future. One could also choose an {\it in}-{\it in} prescription defining the vacuum only at early times, as is mandatory in cosmology, with straightforward changes.
 
Now that we have written the retarded Green's function in terms of a path integral, we can introduce the external sources coupled to $J^\mu$, which effectively gauge the symmetry, and write correlation functions of currents as functional derivatives with respect to the gauge bosons.  Each ${\cal{L}}(\phi_i)$ gets replaced by a ${\cal{L}}\left(\phi_i,A^{(i)}_\mu\right)$ and we can write
\bea
&& G^{\mu\nu}_R(x,y)= \label{eq:GRfunctional}\\  \nn
&&=\frac{i}{Z} \left(\left.\int {\cal{D}}\phi_0\; e^{i\,\int_{\mathbb{R}^d} \di^d x\; {\cal{L}}\left(\phi_0,A^{(0)}_\mu\right)}\; J^\mu\left(\phi_0(x)\right) J^\nu\left(\phi_0(y)\right)\right|_{A_\mu^{(0)}=0}+\right.\\ \nn
&&\quad \int {\cal{D}} \phi(\tilde{\boldsymbol{x}}) \int {\cal{D}} \phi(\tilde{\boldsymbol{y}})\ J^\nu(\phi(y^0,\boldsymbol{y})) J^\mu(\phi(x^0,\boldsymbol{x})) \int_{\phi(\tilde{\boldsymbol{y}})}^{\rm}{\cal{D}}\phi_3 \;e^{ i\,\int_{y^0}^{+\infty} \di^d x\; {\cal{L}}\left(\phi_3,A_\mu^{(3)}\right)}  \times \\  \nn
&&\quad \int_{\phi(\tilde{\boldsymbol{x}})}^{\phi(\tilde{\boldsymbol{y}})}{\cal{D}}\phi_2 \; e^{ i\,\int_{x^0}^{y^0} \di^d x\; {\cal{L}}\left(\phi_2,A_\mu^{(2)}\right)}\ \left.\left. \   \int_{\rm}^{\phi(\tilde{\boldsymbol{y}})}{\cal{D}}\phi_1 \;e^{ i\,\int_{-\infty}^{x^0} \di^d x\; {\cal{L}}\left(\phi_1,A_\mu^{(1)}\right)}\right|_{A_\mu^{(1,2,3)}=0}\right) \;.\eea
Rewriting the currents as derivatives with respect to $A_\mu$ one gets
\bea
&& G^{\mu\nu}_R(x,y)= 
\frac{i}{Z} \left(-\left.\frac{\delta^2}{\delta A_\mu^{(0)}(x)\delta A_\nu^{(0)}(y)}\int {\cal{D}}\phi\; e^{i\,\int_{\mathbb{R}^d} \di^d x\; {\cal{L}}\left(\phi_0,A^{(0)}_\mu\right)}\right|_{A_\mu^{(0)}=0}-\right.\label{eq:GRfunctional2}\\ \nn
&&\quad \frac{\delta^2}{\delta A_\mu^{(1)}(x)\;\delta A_\nu^{(3)}(y)}\int {\cal{D}} \phi(\tilde{\boldsymbol{x}}) \int {\cal{D}} \phi(\tilde{\boldsymbol{y}})\  \int_{\phi(\tilde{\boldsymbol{y}})}^{\rm}{\cal{D}}\phi_3 \;e^{ i\,\int_{y^0}^{+\infty} \di^d x\; {\cal{L}}\left(\phi_3,A_\mu^{(3)}\right)} \  \times \\ \nn
&&\left.\quad\left. \int_{\phi(\tilde{\boldsymbol{x}})}^{\phi(\tilde{\boldsymbol{y}})}{\cal{D}}\phi_2 \; e^{ i\,\int_{x^0}^{y^0} \di^d x\; {\cal{L}}\left(\phi_2,A_\mu^{(2)}\right)}  \int_{\rm}^{\phi(\tilde{\boldsymbol{x}})}{\cal{D}}\phi_1 \;e^{ i\,\int_{-\infty}^{x^0} \di^d x\; {\cal{L}}\left(\phi_1,A_\mu^{(1)}\right)}\right|_{A_\mu^{(1,2,3)}=0}\right)\ .
\eea
This is an expression in the full UV theory. The low-energy EFT can be obtained integrating out the heavy fields: splitting the fields in heavy, $\phi_h$, and light, $\phi_\ell$, one has
\be
e^{i S_{\rm EFT}(\phi_\ell, A_\mu)} = \int {\cal{D}}\phi_h \, e^{i S_{\rm EFT}(\phi_h, \phi_\ell,A_\mu)} \,. 
\ee
The resulting action is gauge invariant and thus contains all the ``minimal'' couplings of $\phi_\ell$ with $A_\mu$ induced by gauging. On top of this, additional local operators depending on  $A_\mu$ and $\phi_\ell$ will be generated when integrating out the heavy fields $\phi_h$.  In particular operators which are quadratic in $A_\mu$ will contribute contact terms to the $\langle JJ\rangle$-correlators, when taking the functional derivatives with respect to $A_\mu$.  It is important to notice that only the time-ordered correlator will contain contact terms: indeed this is the only one in Eq.~\eqref{eq:GRfunctional} that contains second derivatives with respect to the {\em same} $A_\mu$. This is consistent with our discussion above. Notice that one would have missed these contact terms calculating the correlators of the Noether current of the light fields in the EFT.

\paragraph{Contour argument.} We now come to the general argument that gives bounds on the coefficients of operators in theories where boosts are spontaneously broken. In this introductory discussion we set $\kz= \boldsymbol{0}$ for simplicity; we will come back to the general case later. Given a low-energy EFT, characterized by a cutoff $\Lambda$, we can compute $\tilde G^{\mu\nu}(\omega)$ at low energies: schematically for the $00$ component one gets
\be\label{eq:schematic1}
\tilde{G}^{00}(\omega) = \mu^{d-2}\left[c_1 \frac{1}{1-c_s^2 \xi^2}+ \frac{\omega^2}{\Lambda^2}\left(\frac{c_2}{(1-c_s^2\xi^2)^2}+d_1\right)+{\cal{O}}\left(\frac{\omega^4}{\Lambda^4}\right)\right]\ .
\ee 
The denominators come from the propagators of the low-energy degree of freedom, with speed of propagation $c_s$. Here $\mu$ is some overall scale, while $c_1,c_2,\dots,\ d_1,d_2,\ldots$ are coefficients of operators of the low-energy EFT, including terms quadratic in the gauge fields, which give rise to contact terms. The parameter $\xi$, defined in Eq.~\eqref{choicep1} ($0 \le \xi <1$), allows to explore the region of analiticity of $\tilde G$. 

Now the argument is similar to the $S$-matrix one (see Fig.~\ref{fig:contour}). One can select the coefficient of a given power of $\omega$ in~\eqref{eq:schematic1} using the residue theorem
\be
\ointctrclockwise \di \omega\frac{\tilde{G}^{00}(\omega)}{\omega^3}=2\pi i \left(\frac{c_2}{(1-c_s^2\xi^2)^2}+d_1\right)\frac{\mu^{d-2}}{\Lambda^2} \ .
\ee
The contour of integration can then be moved far away from the origin as in Fig.~\ref{fig:contour}. The contribution of the circle at infinity can be estimated using the CFT correlator, Eq.~\eqref{eq:JJCFT}, which is appropriate for the $|\omega| \to \infty$ limit.\footnote{One might be worried that the CFT limit is reached for large {\em real} momenta, while the contour at infinity needs large $|\omega|$ in the complex plane. Let us show that the knowledge of the UV limit of the Green's functions in position space for real arguments gives us control in Fourier space for complex $\omega$. Let us consider
\begin{equation} 
	\tilde{G}_{R}(\omega,\omega \boldsymbol{\xi}) = \int_{\mathbb{R}^d} \di^d x \, e^{-i p \cdot x} G_{R}(x) \,,
\end{equation}
where we dropped $\boldsymbol{k}_0$, which is negligible for large $|\omega|$, and suppressed for simplicity the indices of the Green's function. (We focus on the retarded Green's function in the upper half plane, the same holds for $G_A$ in the lower half.)
We want to study the limit of large $|\omega|$ for a fixed direction in the complex plane, i.e.~the limit $\lambda \to +\infty$ ($\lambda$ is real and we assume $\lambda \ge 1$) of 
\begin{equation} \label{eq:lambdalimit}
	\tilde{G}_{R}(\lambda \omega,\lambda\omega \boldsymbol{\xi}) = \int_{\mathbb{R}^d} \di^d x \, e^{-i \lambda p \cdot x} G_{R}(x) = \lambda^{-d}\int_{\mathbb{R}^d} \di^d x \, e^{-i p \cdot x} G_{R}\left(\frac{x}{\lambda}\right)\,,
\end{equation}
where in the last step we redefined the variables of integration. We are assuming that the Green's functions are polynomially bounded in position space: $|G_R(x)| \le A |x|^n$ for $|x|$ above a certain value and for suitable $A >0$ and positive integer $n$. This implies that the integral above converges exponentially for $p^\textsf{Im} \in \text{FLC}$ and one can neglect, for any $\lambda \ge 1$, large values of $|x|$. More precisely, once one fixes a small $\varepsilon$ one can find a sufficiently large, $\lambda$-independent,  $x_\Lambda$ such that 
\be
\left|\int_{|x| > x_\Lambda} \di^d x \, e^{-i p \cdot x} G_{R}\left(\frac{x}{\lambda}\right) \right| \le \int_{|x| > x_\Lambda} \di^d x \, \left|e^{-i p \cdot x}\right| \left|G_{R}\left(\frac{x}{\lambda}\right) \right| \le \int_{|x| > x_\Lambda} \di^d x \, e^{p^\textsf{Im} \cdot x} A |x|^n < \varepsilon \;.
\ee
In the compact region $|x| < x_\Lambda$, one has $G_R(x/\lambda) \to G_R^\text{CFT}(x/\lambda)$ for $\lambda \to + \infty$ and the convergence is uniform in $x$. Given the uniform convergence, in Eq.~\eqref{eq:lambdalimit} one can exchange the $\lambda$ limit with the integral and conclude that the result converges to the CFT result for all complex $\omega$ in the upper half plane.} For example in $d=3$, $\tilde{G}^{00}(\omega)\sim \omega$ in this limit and the contribution from the circle at infinity is negligible.  Thus the integral around the origin is equal to the integral along the cuts, which is $i\times ({\rm positive})$. We therefore conclude:
\be\label{eq:schematicres}
\frac{c_2}{(1-c_s^2\xi^2)^2}+d_1\geq 0\ .
\ee
The most general inequalities can be obtained varying $\xi$ in the interval $0 \le \xi <1$ and contracting $\tilde{G}^{\mu\nu}$ with a generic vector $V^\mu$.

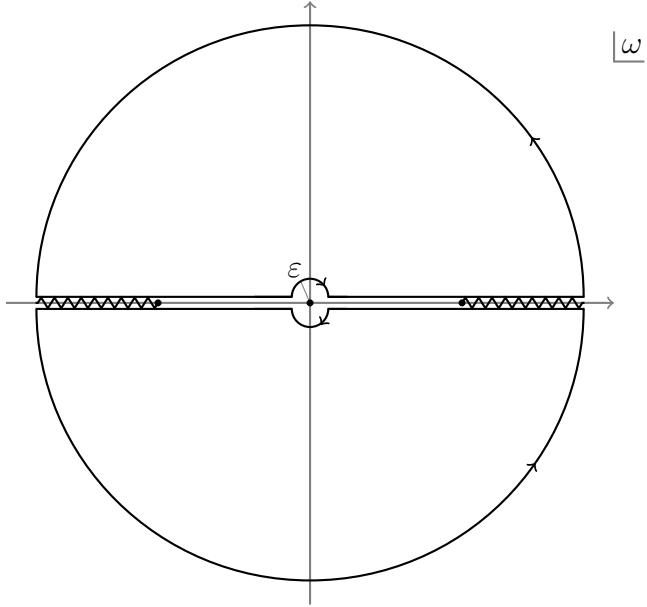
\begin{figure}[h!]
\centering
\begin{tikzpicture}[thick,scale=0.8]

  \draw[->,gray] (-5,0) -- (5,0);
  \draw [->,gray] (0,-5) -- (0,5);

  \draw[gray, thick] (5,4) -- (5.5,4);
  \draw[gray, thick] (5,3.987) -- (5,4.5);
  \node at (5.3,4.25) {$\omega$};
        
        \draw[black,
        decoration={markings, mark=at position 0.2 with {\arrow{>}}},
        postaction={decorate}] (4.5,0.1) arc
    [
        start angle=0,
        end angle=180,
        x radius=4.5cm,
        y radius =4.5cm
    ];
    
        \draw[black,
        decoration={markings, mark=at position 0.2 with {\arrow{<}}},
        postaction={decorate}] (4.5,-0.1) arc
    [
        start angle=0,
        end angle=-180,
        x radius=4.5cm,
        y radius =4.5cm
    ];
    
    \draw[gray,thin] (-1.2+1.2,0) -- (-1.35+1.2,0.35);
    \node at (-1.4+1.15,0.55) {$\varepsilon$};
    \draw[black] (-4.514,0.1) -- (-0.3,0.1);
    \draw[black] (-4.514,-0.1) -- (-0.3,-0.1);
    \draw[black] (-0.914,0.1) -- (-0.286,0.1);
    \draw[black] (-0.914,-0.1) -- (-0.286,-0.1);
    \draw[black] (0.286,0.1) -- (0.614,0.1);
    \draw[black] (0.286,-0.1) -- (0.614,-0.1);
    
    \draw[black,decoration={markings, mark=at position 0.8 with {\arrow{>}}},
        postaction={decorate}] (-0.3,0.1) arc
    [
        start angle=180,
        end angle=0,
        x radius=0.3cm,
        y radius=0.3cm
    ];
    
    \draw[black,decoration={markings, mark=at position 0.8 with {\arrow{<}}},
        postaction={decorate}] (-0.3,-0.1) arc
    [
        start angle=-180,
        end angle=0,
        x radius=0.3cm,
        y radius=0.3cm
    ];

    \draw[black] (0.3,0.1) -- (4.514,0.1);
    \draw[black] (0.3,-0.1) -- (4.514,-0.1);
    \draw[black,fill=black] (0,0) circle (0.04cm);
    
    \draw[decorate,decoration={zigzag, segment length=5, amplitude=2}] (-2.5,0) -- (-4.514,0);
    \draw[black,fill=black] (2.5,0,0) circle (0.04cm);
    \draw[decorate,decoration={zigzag, segment length=5, amplitude=2}] (2.5,0) -- (4.514,0);
    \draw[black,fill=black] (-2.5,0,0) circle (0.04cm);
 \end{tikzpicture}
\caption{Different contours that give the same result as in Fig.~\ref{fig:contour}. This choice is suitable when a cut is running all along the real axis.} \label{fig:doublecontour}
\end{figure}

\paragraph{Getting rid of the mass-gap assumption.} The contour argument above assumes that one can stick to the tree-level approximation in the EFT, so that at low energy one has only poles. In general, however, in the absence of a mass gap, loops of the low-energy excitation will open a cut in the $\omega$-plane all along the real axis. In this case one can use the two contours of Fig.~\ref{fig:doublecontour}. One gets the same result as with the contour of Fig.~\ref{fig:contour}, but now this can be applied even in the presence of a cut running all along the real axis. In Appendix \ref{k0sec} we will see that the positivity arguments can also be derived remaining in the upper half plane, i.e.~using only the upper contour of Fig.~\ref{fig:doublecontour}

In the remainder of the paper we will apply the above ideas to the interesting case of an EFT where boosts are non-linearly realized: a CFT at finite chemical potential.

\section{Conformal superfluids \label{superfluidsec}}
CFT's are of fundamental interest in physics and any technique to get extra information about them is precious. A way to understand analytically the general properties of operators at large charge $Q$ was developed in \cite{Hellerman:2015nra, Monin:2016jmo} and in many subsequent papers, for a review see \cite{Gaume:2020bmp}. By the state/operator correspondence, a charged operator is mapped into a charged state. Generically this charge induces the spontaneous breaking of the $U(1)$ symmetry (for simplicity we focus on a $U(1)$). We are therefore lead to study CFT's at finite chemical potential $\mu$: one can think about this as a state that evolves in time around the $U(1)$. Lorentz invariance is thus spontaneously broken and this connects to the topic of the present paper. The breaking of the symmetry leads to a Goldstone boson: this is the only degree of freedom at energies much smaller than $\mu$. As we will discuss below the action for this Goldstone is very constrained by the symmetry breaking pattern. The original symmetry, $SO(d,2) \times U(1)$, is broken to rotations and spacetime translations, with the time translation being a diagonal combination of the original time translation and a $U(1)$ rotation.
One can think about this symmetry-breaking pattern considering a charged scalar $\chi$ evolving linearly in time 
\be \label{chipi}
\chi(x) = \mu t + \pi(x) \,,
\ee
where $\pi$ is the Goldstone boson.
The most general action can be obtained through a coset construction \cite{Monin:2016jmo} or, maybe more simply, using an effective metric as we discuss in detail in Appendix \ref{EFToper}.

This EFT contains all the information about the large $Q$ sector of the theory, and the derivative expansion can be identified with the large charge expansion. The simplest object one can calculate is the energy of the system as a function of $Q$, which gives the lowest scaling dimension of operators of charge $Q$. In general CFT correlation functions involving at least two operators with large charge can be calculated using the EFT.\footnote{If one is interested in operators which are not close to the bottom of the spectrum, one has to consider a very excited quasi-thermal state of the superfluid: in this case dissipative hydrodynamics is a better description \cite{Delacretaz:2020nit}.} All the details about the specific CFT at hand are encoded in the coefficients of the operators of the EFT: our positivity constraints will thus carve out a region of possible CFT data. 

Besides its intrinsic interest, this EFT is the perfect example to apply our methods to. The UV CFT, which is a crucial step for our argument, is here broken only by $\mu$, which is also responsible for the breaking of the Lorentz symmetry. In some sense this is the minimal setup where our arguments can be tested and applied. Notice that the EFT only makes sense as Lorentz-breaking: the cut-off of the theory is $\mu$, the only scale in the problem, and it goes to zero as one tries to switch off the Lorentz-breaking.

Although our arguments are general we are going from now on to work in $d=3$. One reason is that most of the CFT literature concentrates on this example, since $d=3$ CFT's are the key objects to study second-order phase transitions. Another reason is that in $d=3$ we are able to constrain the NLO operators of the theory, while in $d=4$ the contour at infinity would not converge and one would have to go to higher order.

\subsection{$\langle JJ \rangle$ calculation \label{JJsec}}
The EFT Lagrangian for $\pi$ coupled to $A_\mu$ in $d=3$ reads, to NLO in derivatives (see \cite{Cuomo:2021qws} and Appendix \ref{EFToper}),
\begin{align} \label{gaugeL}
	\mathcal{L} &= \frac{c_1}{6} |\nabla \chi|^3 - 2 c_2 \frac{(\partial |\nabla \chi|)^2}{|\nabla \chi|} + c_3 \left( 2 \frac{\left( \nabla^\mu \chi \partial_\mu |\nabla \chi| \right)^2}{|\nabla \chi|^3} + \partial_\mu \left( \frac{\nabla^\mu \chi \nabla^\nu \chi}{|\nabla \chi|^2} \right) \partial_\nu |\nabla \chi| \right) \notag \\
	&- \frac{b}{4} \frac{F_{\mu \nu} F^{\mu \nu}}{|\nabla \chi|} +  \frac{d}{2} \frac{F^{\mu}_{\;\; i} F^{\nu i}}{|\nabla \chi|^3} \nabla_\mu\chi \nabla_\nu\chi \,,
\end{align}
where $\chi$ is expressed in terms of $\pi$ as in~\eqref{chipi} and
\begin{align}
	\nabla_\mu \chi &\equiv \partial_\mu \chi - A_\mu \,, \label{covariantchi} \\
	|v| &\equiv \sqrt{-v_\mu v^\mu} \,.
\end{align}
The leading order term, the one proportional to $c_1$, makes sense only when $\chi$ is expanded around the time-dependent background of Eq.~\eqref{chipi}: the EFT cutoff is set by the only scale in the problem, $\mu$, so the EFT loses sense if one tries to extrapolate to the Lorentz-invariant vacuum $\mu = 0$. Around the background $\eqref{chipi}$, $\pi$ has a speed of propagation $c_s^2 =1/2$, fixed by the conformal symmetry (it would be $c_s^2 = 1/3$ in $d = 4$).  

Notice the gauge symmetry $\pi(x) \rightarrow \pi(x) + \Lambda(x), A_\mu(x) \rightarrow A_\mu(x) + \partial_\mu \Lambda(x)$ since $\chi$ only appears in the combination~\eqref{covariantchi}, and also the two quadratic kinetic terms for $A_\mu$ which are compatible with the spontaneous breaking of Lorentz invariance. We argued in \S\ref{sec:setup} that these terms produce contact terms in the two-point function of $J^\mu$ that one may not neglect since they will be generated when integrating out the heavy fields. As such they appear with the unknown coefficients $b$ and $d$ in the EFT.\footnote{If we were to make $A_\mu$ dynamical these operators would encode the modification of the ``photon'' propagator inside the material, inducing in particular a modification of the speed of light.} These operators were not discussed in the CFT literature (see for instance \cite{Monin:2016jmo,Cuomo:2021qws}) because they did not study correlators involving the $U(1)$ current. Finally, notice one needs $c_1 >0$ to have a healthy kinetic term for $\pi$. We will now constrain the other coefficients $c_2, c_3, b$ and $d$ by using the methods outlined above.

Expanding to second order in $\pi,A$ we have, after integration by parts,
\begin{align}
	\mathcal{L}_{(2)} &= \frac{c_1 \mu^3}{6} + \frac{\mu c_1}{2} \left[ (\dot{\pi} + A^0)^2 - \frac{1}{2} \left( \partial_i \pi - A_i \right)^2 + \mu \left( \dot{\pi} + A^0 \right) \right] + \frac{2 c_2}{\mu} \left[ - \pi \square \ddot{\pi} + 2 A^0 \square \dot{\pi} + A^0 \square A^0 \right] \notag \\
	&+\frac{2 c_3}{\mu} \left[ -\pi \square \ddot{\pi} + 2 A^0 \square_{c_s} \dot{\pi} - A^i \partial_i \ddot{\pi} + (\dot{A}^0)^2 + \dot{A}^0 \partial_i A^i \right] + \notag \\ & + \frac{(b+d)}{2\mu} \left[ (\partial_i A^0)^2 + (\partial_0 A_i)^2 + 2 \dot A^0 (\partial_i A_i)\right] 
	- \frac{b}{4 \mu} \left( \partial_i A_j - \partial_j A_i \right)^2  \label{eq:quadaction}\,,
\end{align}
where $\square \equiv \partial_\mu \partial^\mu$ and $\square_{c_s} \equiv -\partial^2_t + c_s^2 \partial_i \partial^i$ with $c_s^2 = 1/2$. The conserved Noether current $J^\mu_N$  is the one associated to the symmetry $\chi \rightarrow \chi + c$ of~\eqref{gaugeL} and reads, to first order in $\pi$,
\begin{align}
	J_N^0 &= -\frac{\mu^2 c_1}{2} -\mu c_1 \dot{\pi} - \frac{4 c_2}{\mu} \square \dot{\pi} - \frac{4 c_3}{\mu} \square_{c_s} \dot{\pi} \,, \\
	J_N^i &= \frac{\mu c_1}{2} \partial_i \pi - \frac{2 c_3}{\mu} \partial_i \ddot{\pi} \,,
\end{align}
which may be derived from the Noether procedure writing $\mathcal{L} = \mathcal{L}(A=0) + A_\mu J_N^\mu + \mathcal{O}(A^2)$.

To calculate the two-point function of $J_N^\mu$ in the EFT we will require the propagator of $\pi$. We obtain it from the quadratic action with $A = 0$,
\begin{align}
	\mathcal{L}_{(2),A=0} &= \frac{\mu c_1}{2} \pi \square_{c_s} \pi - \frac{2 (c_2+c_3)}{\mu} \pi \square \ddot{\pi} \notag \\
	&= \frac{1}{2} \dot{\pi}_c^2 - \frac{1}{4} \left( \partial_i \pi_c \right)^2 + \frac{2 (c_2+c_3)}{\mu^2 c_1} \left( \ddot{\pi}_c^2 - \left( \partial_i \dot{\pi}_c \right)^2 \right) \,,
\end{align}
where we have defined a canonical field $\pi_c \equiv \sqrt{\mu c_1} \pi$. In Fourier space this is
\begin{equation}
	\mathcal{L}_{(2),A=0}(k) = \frac{1}{2} \tilde{\pi}_c(-k) \left( \omega^2 - c_s^2 \boldsymbol{k}^2 + \frac{4 (c_2+c_3)}{\mu^2 c_1} \omega^2 \left( \omega^2 - \boldsymbol{k}^2 \right) \right) \tilde{\pi}_c(k) \,.
\end{equation}
(Recall our Fourier convention $f(x) = (2\pi)^{-3} \int \di^3 k \, e^{ik \cdot x} \tilde{f}(k)$.) The $\tilde{\pi} \tilde{\pi}$ propagator is then, to first subleading order in derivatives,
\begin{equation}
	\langle \tilde{\pi}_c(-k) \tilde{\pi}_c(k) \rangle = \frac{i}{\omega^2 - c_s^2 \boldsymbol{k}^2} \left( 1 - \frac{4 (c_2+c_3)}{\mu^2 c_1} \frac{\omega^2 \left( \omega^2 - \boldsymbol{k}^2 \right)}{\omega^2 - c_s^2 \boldsymbol{k}^2} \right) \,.
\end{equation}
Here and in the following we are removing a $(2\pi)^3\delta^{(3)}(k+k')$ of conservation of energy and momentum, and when we suppress the external states we will always consider expectation values in the interacting vacuum. We did not specify the prescription for the poles: the analytic structure is the one discussed in \S\ref{sec:setup} corresponding to retarded Green's functions as we approach from the upper half $\omega$-plane and advanced approaching from below.

Let us now proceed to calculate the current-current correlator. The retarded Green's function is written as the difference of the T-ordered and the unordered correlator in Eq.~\eqref{eq:green_time}. Let us start with the T-ordered piece, whose functional representation is the first line of Eq.~\eqref{eq:GRfunctional2}.  This piece, as discussed in \S\ref{sec:setup}, will not only contain the T-ordered correlator of the Noether currents, but also contact terms, which are crucial in order to match with the UV calculation. These contact terms are given in Eq.~\eqref{eq:GRfunctional2} when one takes the derivatives to act twice on the Lagrangian: 
\be\label{eq:coincident}
\frac{1}{Z} \int {\cal{D}}\phi\; e^{i\,\int_{\mathbb{R}^3} \di^3 x\; {\cal{L}}\left(\phi_0,A^{(0)}_\mu\right)} \left.\frac{\delta^2 {\cal{L}}\left(\phi_0,A^{(0)}_\mu\right)}{\delta A_\mu^{(0)}(x)\delta A_\nu^{(0)}(x)}\right|_{A_\mu^{(0)}=0} \;.
\ee
(When the functional derivatives act on two different ${\cal{L}}$'s, one gets the correlation functions of the Noether currents.)
Equation~\eqref{eq:coincident}, at tree level, simply takes the terms in the action Eq.~\eqref{eq:quadaction} which are quadratic in $A^\mu$ and contributes to $G_R^{\mu\nu}$ contact terms in real space (derivatives of the delta function), i.e.~polynomials in $\omega$ and $\boldsymbol{k}$ in Fourier space.  Now let's consider the unordered correlator, the second line of Eq.~\eqref{eq:GRfunctional2}. We discussed above that this does not contribute to contact terms. In general, this is a quite complicated object: since it is not T-ordered, one should develop the proper Feynman rules. However, in this paper we will stick to tree-level two-point function calculations in the EFT and one expects the unordered term to combine with T-ordered correlator of currents to give the retarded Green's function of currents, see Eq.~\eqref{eq:green_time}. Indeed one can see that the unordered term just changes the prescription of the $\omega < 0$ poles to make them retarded. Eq.~\eqref{eq:KLunordered} shows that the unordered correlator at a given $\omega$ and $\boldsymbol{k}$ just receives contribution from states of the theory with same $\omega$ and $\boldsymbol{k}$: at tree level,  this implies that its Fourier transform is localized on the poles of the propagator. Indeed the sum over $n$ in Eq.~\eqref{eq:KLunordered} is actually an integral over the one-particle states $\di \tilde{\boldsymbol{k}}/(2 \pi)^{d-1} \cdot 1/(2 \omega(k'))$: this integral eats the spatial delta function and gives
\be
 i\braket{0| J^\nu(-\omega, -\boldsymbol{k}) J^\mu(\omega, \boldsymbol{k}) |0} = i(2\pi) \frac{1}{2 \omega(k)}\delta(\omega + \omega(k)) \braket{0|J^\nu(0)|\omega(k),\boldsymbol{k}} \braket{\omega(k),\boldsymbol{k}|J^\mu(0)|0}  \,,
\ee
where $\omega(k)$ is the dispersion relation of the Goldstone (including the corrections due to higher-dimension operators).
Notice that this contribution is only for $\omega <0$: one can check that this term changes the prescription of the poles, exactly in the same way it does for the standard relativistic propagator of a massive scalar field.

The structure of $\langle J^\mu(-k) J^\nu(k)\rangle$ is severely constrained by current conservation,
\be
k_\mu \langle J^\mu(-k) J^\nu(k)\rangle =0 \ .
\ee
This equality is exact, without contact terms on the RHS, given our definition of the correlator: we show this precisely in Appendix \ref{app:conservation}.
In the absence of Lorentz invariance one has two possible tensor structures that guarantee conservation,
\begin{align}
	i \langle J^\mu(-k) J^\nu(k)\rangle = \textsf{A} \left( k^\mu k^\nu - \eta^{\mu \nu} k^2 \right) + \textsf{B} \left( k^i k^j - \delta^{ij} \boldsymbol{k}^2 \right) \,,\label{JmuJnu}
\end{align}
where $\textsf{A}$ and $\textsf{B}$ are general functions of $\omega$ and $|\boldsymbol{k}|$.

In our EFT, calculating the Noether current correlator\footnote{Notice that the constant term in $J^0_N$ drops out from the Green's function, since it involves a commutator.} and adding the contact terms one gets
\begin{align}
	\textsf{A} &= -\frac{\mu c_1}{2 \left( \omega^2 - c_s^2 \boldsymbol{k}^2 \right)} + \frac{c_2}{\mu} \frac{\left( \omega^2 - \boldsymbol{k}^2 \right) \boldsymbol{k}^2}{\left( \omega^2 - c_s^2 \boldsymbol{k}^2 \right)^2} - \frac{c_3}{\mu} \frac{\omega^2 \boldsymbol{k}^2}{\left( \omega^2 - c_s^2 \boldsymbol{k}^2 \right)^2} + \frac{b}{\mu} + \frac{d}{\mu} \,, \label{JJAEFT} \\
	\textsf{B} &= \frac{\mu c_1}{4 \left( \omega^2 - c_s^2 \boldsymbol{k}^2 \right)} + \frac{c_2}{\mu} \frac{\left( \omega^2 - \boldsymbol{k}^2 \right)^2}{\left( \omega^2 - c_s^2 \boldsymbol{k}^2 \right)^2} - \frac{c_3}{\mu} \frac{\omega^2(\omega^2-\boldsymbol{k}^2)}{\left( \omega^2 - c_s^2 \boldsymbol{k}^2 \right)^2} - \frac{d}{\mu} \,. \label{JJBEFT}
\end{align}
The prescription for the poles is retarded for $\omega^\textsf{Im} >0 $ and advanced for $\omega^\textsf{Im} <0$.

\subsection{Positivity bounds from $\langle JJ \rangle$ \label{JJpossec}}
To derive positivity bounds on $c_2, c_3, b$ and $d$ we follow the general logic outlined in \S\ref{sec:setup} (see also Appendix \ref{k0sec}). For the time being, we focus on the tree-level approximation in the EFT and we will discuss loops in \S\ref{sec:loops}. Consider the function
\begin{equation}
	\tilde{f}(\omega) = \tilde{G}^{\mu \nu}(k) V_\mu(k) V_\nu(k) \Big|_{k = (\omega,\boldsymbol{k}_0 + \omega \boldsymbol{\xi})} \,, \label{ftreeEFT}
\end{equation}
where
\begin{equation}
	\tilde{G}^{\mu \nu}(k) = i \langle J^\mu(-k) J^\nu(k)\rangle \,.
\end{equation}
Its EFT low-energy approximation is given by Eqns.~\eqref{JmuJnu},~\eqref{JJAEFT} and~\eqref{JJBEFT}. The vector $V(k = (\omega, \boldsymbol{k}_0 + \omega \boldsymbol{\xi})) \equiv V(\omega)$ has components which are arbitrary polynomials in $\omega$. Initially we set $\boldsymbol{k}_0 = \boldsymbol{0}$, but turning on this parameter does not produce any new bounds as we illustrate explicitly at the end of this section. At each $\omega$ we may expand $V(\omega)$ as a sum of three terms, one parallel to $k$ and two others orthogonal to it, with arbitrary coefficients as long as the result is a polynomial in $\omega$. When $\boldsymbol{k}_0 = \boldsymbol{0}$ these basis vectors are $\omega$-independent when normalized:
\begin{align}
	\hat{K} &  = \frac{(1,\boldsymbol{\xi})}{\sqrt{1-\xi^2}} \,, \label{Khat} \\
	\hat{E} &= \frac{(\xi,\hat{\boldsymbol{\xi}})}{\sqrt{1-\xi^2}} \,, \label{Ehat} \\
	\hat{F} &= (0,\hat{\boldsymbol{f}}) \,, \label{Fhat}
\end{align}
where hats denote unit vectors, so $\hat{\boldsymbol{\xi}} \cdot \hat{\boldsymbol{\xi}} = 1 = \hat{\boldsymbol{f}} \cdot \hat{\boldsymbol{f}}$, $\hat{\boldsymbol{\xi}} \cdot \hat{\boldsymbol{f}} = 0$. Also $\hat{K} \cdot \hat{K} = -1, \hat{E} \cdot \hat{E} = 1 = \hat{F} \cdot \hat{F}$ and $\hat{K} \cdot \hat{E} = \hat{K} \cdot \hat{F} = \hat{E} \cdot \hat{F} = 0$. So we write
\begin{equation} \label{Vexpansion}
	V(\omega) = \alpha(\omega) \hat{K} + \beta(\omega) \hat{E} + \gamma(\omega) \hat{F} \,,
\end{equation}
where $\alpha, \beta, \gamma$ are arbitrary polynomials of $\omega$. This expansion is useful because it is immediate from~\eqref{JmuJnu} that $\alpha(\omega)$ will not appear in the sum~\eqref{ftreeEFT}. This property of the expansion persists when $\boldsymbol{k}_0 \neq \boldsymbol{0}$.

An essential step in the positivity logic is that we must be able to neglect the contribution from complex infinity to the contour integral. The full function $\tilde{f}(\omega) = \tilde{G}^{\mu \nu}(\omega) V_\mu(\omega) V_\nu(\omega)$ behaves as $\omega \times \omega^{2N}$ as $|\omega| \rightarrow \infty$ (from the CFT result~\eqref{eq:JJCFT}), where $N$ is the highest degree among $\beta(\omega)$ and $\gamma(\omega)$. To neglect the arc at infinity, we must divide by at least $\ell = 3 + 2N$ powers of $\omega$. Near $\omega = 0$, however, $\langle JJ \rangle$ behaves schematically as $\mu + \omega^2/\mu + \omega^4/\mu^3 + \cdots$. Since in~\eqref{JmuJnu} we have only computed $\langle JJ \rangle$ to order $\omega^2$, we must take $N = 0$ to pick out those terms.

\begin{figure}[h!]
\centering
\includegraphics[width=0.6\textwidth]{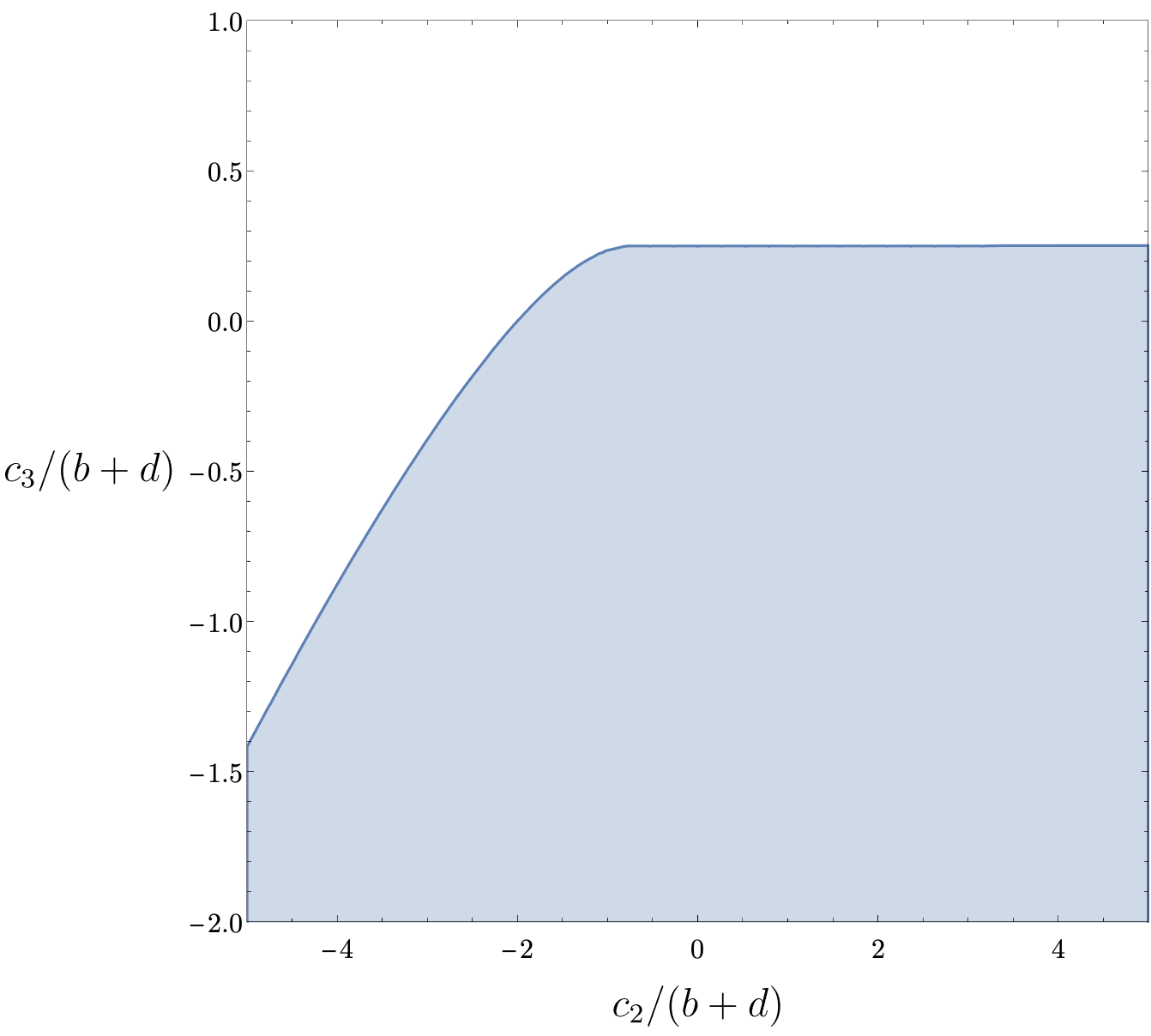}
\caption{\small The constraints~\eqref{ineq48} (the blue region is allowed). There are also the constraints $d \geq 0$ and $b+d\geq0$. $c_1 >0$ is required to have a healthy kinetic term for fluctuations.}
\label{c2c3fig}
\end{figure}

So in our case $\alpha,\beta,\gamma$ in~\eqref{Vexpansion} are just numbers, and a straightforward calculation yields
\begin{align}
	\tilde{f}(\omega) &= \textsf{A} \omega^2 (1-\xi^2)(\beta^2 + \gamma^2) - \textsf{B} \xi^2 \omega^2 \gamma^2 \notag \\
	&= -\frac{\mu c_1}{2} \frac{1}{1-\xi^2/2} \left[ (1-\xi^2) \beta^2 + (1-\xi^2/2) \gamma^2 \right] + \frac{c_2}{\mu} \omega^2 \frac{\xi^2(1-\xi^2)^2}{(1-\xi^2/2)^2} \beta^2 \notag \\&- \frac{c_3}{\mu} \omega^2 \frac{\xi^2(1-\xi^2)}{(1-\xi^2/2)^2} \beta^2 + \frac{b}{\mu} \omega^2 (1-\xi^2)(\beta^2 + \gamma^2) + \frac{d}{\mu} \omega^2 \left[ (1-\xi^2)\beta^2 + \gamma^2 \right] \,.
\end{align}
We now follow the contour argument of \S\ref{sec:setup} and consider a contour around the origin (see Fig.~\ref{fig:contour})
\be
\ointctrclockwise \di \omega\frac{\tilde{f}(\omega)}{\omega^3} = i\pi \tilde{f}''(0) \;.
\ee
This contour can be deformed to the integral around the cut and the circle at infinity. The circle at infinity is negligible as one can see using the CFT result Eq.~\eqref{eq:JJCFT} in $d=3$. The integral around the cut is $i \times \text{(positive)}$ as shown in \S\ref{sec:setup}. So
\begin{align}
	\tilde{f}''(0) \geq 0 \,.
\end{align}
This reads
\begin{equation} \label{eq47}
	c_2 \frac{\xi^2(1-\xi^2)}{(1-\xi^2/2)^2} \beta^2 - c_3 \frac{\xi^2}{(1-\xi^2/2)^2} \beta^2 + b (\beta^2 + \gamma^2) + d \left( \beta^2 + \frac{\gamma^2}{1-\xi^2} \right) \geq 0
\end{equation}
for all choices $\xi \in [0,1)$ and $\beta,\gamma$. These bounds may be reformulated as follows: first, letting $\xi \rightarrow 1$ with $\gamma \neq 0$ we obtain $\boxed{d \geq 0}$, while letting $\xi \rightarrow 0$ we get $\boxed{b+d \geq 0}$. Putting the terms proportional to $\gamma^2$ on the RHS of the inequality, we observe that the most stringent bound is obtained at $\gamma = 0$ (because the RHS is negative, and zero when $\gamma = 0$). For $\gamma = 0$ we have
\begin{equation} \label{ineq48}
	\boxed{\frac{c_2}{b+d} (1-\xi^2) - \frac{c_3}{b+d} \geq -\frac{(1-\xi^2/2)^2}{\xi^2} \,.}
\end{equation}
Again these constraints hold for all $\xi \in [0,1)$, and they are plotted in Fig.~\ref{c2c3fig}.
For $\tilde{c}_2 \leq -3/4, \tilde{c}_3 \leq 1/4$ the boundary curve is given by
\begin{equation}
	\tilde{c}_3 = \tilde{c}_2 - 1 + \sqrt{1-4\tilde{c}_2} \,,
\end{equation}
where $\tilde{c}_{2,3} \equiv c_{2,3}/(b+d)$, while for $\tilde{c}_2 \geq -3/4$ the boundary curve is the horizontal line $\tilde{c}_3 = 1/4$.

In terms of CFT applications, the coefficient $c_2$ is the most interesting, since it controls the NLO correction to the dimension of the lowest operator of charge $Q$ in the limit of large charge (see e.g.~\cite{Monin:2016jmo}). The bounds above, unfortunately, do not say something useful on this coefficient alone. 

\paragraph{$\boldsymbol{k}_0 \neq \boldsymbol{0}$.} Let us now explore what happens if one takes $\boldsymbol{k} = \boldsymbol{k}_0 + \omega \boldsymbol{\xi}$ with $\boldsymbol{k}_0 \neq \boldsymbol{0}$ an $\omega$-independent vector, as in Eq.~\eqref{ftreeEFT}.

 We consider polarization vectors of the form $V = \alpha K + \beta E + \gamma F$ with
\begin{align}
	K &= k = (\omega,\boldsymbol{k}) \,, \label{Kmu} \\
	E &= (\boldsymbol{k}^2, \omega \boldsymbol{k}) \,, \label{Emu} \\
	F &= (0,- \omega k_2,\omega k_1) \,, \label{Fmu}
\end{align}
where $\boldsymbol{k} = (k_1,k_2)$. In this way, $K \cdot E = K \cdot F = E \cdot F = 0$.  These vectors are linearly independent (except for the the particular values $\omega = \pm|\boldsymbol{k}|$).

A short calculation yields
\begin{align} \label{KKVV}
	\tilde{f}(\omega) &= \textsf{A}(\omega^2-\boldsymbol{k}^2)\boldsymbol{k}^2 \left( (\omega^2-\boldsymbol{k}^2) \beta^2 + \omega^2 \gamma^2 \right) - \textsf{B}\boldsymbol{k}^4 \omega^2 \gamma^2 \,.
\end{align}
Now we divide by $\omega^3 \times \omega^4$ and integrate along a circle centered at the origin. As opposed to the case $\boldsymbol{k}_0 = \boldsymbol{0}$,~\eqref{KKVV} contains two new poles, apart from the old one at $\omega = 0$, which are located at the $\omega$'s for which $\omega^2 = c_s^2 \boldsymbol{k}^2(\omega)$. That is, at
\begin{align}
	\omega_\pm = \frac{k_0}{2(1-\xi^2/2)} \left( \xi \cos \theta \pm \sqrt{2 - \xi^2 \sin^2 \theta} \right) \,,
\end{align}
where $\theta$ is the angle between $\boldsymbol{k}_0$ and $\boldsymbol{\xi}$. We consider a contour which encircles all three of these low-energy poles, so that the resulting integral is positive, equal to the high-energy contributions. Summing the three contributons, we\footnote{Read: Mathematica.} obtain exactly Eq.~\eqref{eq47}, which is the result of the $\boldsymbol{k}_0 = \boldsymbol{0}$ computation. So $\boldsymbol{k}_0 \neq \boldsymbol{0}$ brings no new constraints.

In hindsight, one can argue that $\boldsymbol{k}_0 \neq \boldsymbol{0}$ cannot give new constraints, at least at the order to which we are working. In order to trust the EFT calculation, we need to take $k_0/\mu\ll 1$. Cauchy's theorem relates the integral in the EFT region to the integral in the UV region. This second integral gets contributions from large energy and momenta, of order $\mu$ or higher. This means that taking $k_0\neq 0$ gives a small correction, suppressed by powers of $k_0/\mu$. Since the UV integral remains almost the same, also the low-energy one must be approximately the same (and it must have the same expression in terms of the EFT coefficients, otherwise we would have two EFT expressions that are forced to be, approximately, equal without symmetry reasons). We conclude that $\boldsymbol{k}_0 \neq \boldsymbol{0}$ gives only tiny corrections to the bounds. The presence of the scale $\boldsymbol{k}_0$ will make operators with a different number of derivatives contribute: so $\boldsymbol{k}_0 \neq \boldsymbol{0}$ will be relevant when one wants to write bounds on operators of different orders, analogously to what was done for example in \cite{Bellazzini:2020cot} in the Lorentz-invariant case. With the same logic one can argue that having a non-constant polynomial for the vector $V^\mu$ does not change the bounds at leading order: one is dominated by the leading monomial and the other terms give subleading corrections like in the case of $\boldsymbol{k}_0 \neq \boldsymbol{0}$.

\subsection{$\langle TT \rangle$ calculation \label{TTsec}}
We now apply the same techniques to the stress-energy tensor. By dimensional analysis the CFT correlator for the stress-energy tensor will contain two more powers of $\omega$ compared to the one involving currents. Therefore to guarantee convergence of the contour at infinity, one has to divide by two more powers of $\omega$ and this means one has to focus on the EFT operators with two more derivatives. 
To calculate the two-point function of the stress-energy tensor we couple $\chi$ to a non-dynamical metric $g_{\mu \nu}$, and write the most general action which non-linearly realizes conformal symmetry. As we discuss in detail in Appendix \ref{EFToper}, this is most easily expressed in terms of the Weyl-invariant combination $\hat{g}_{\mu \nu} = g_{\mu \nu} |g^{\alpha \beta} \partial_\alpha \chi \partial_\beta \chi|$.  At NNLO in derivatives and focussing only on operators that start quadratic in the fluctuations (the only ones that contribute to the two-point function of the stress-energy tensor) one has three new operators
\begin{equation} \label{EFThataction}
	S = \int \di^3 x \sqrt{-\hat{g}} \left( \frac{c_1}{6} - c_2 \hat{R} + c_3 \hat{R}^{\mu \nu} \hat{\partial}_\mu \chi \hat{\partial}_\nu \chi + c_4 \hat R^2 + c_5 \hat R_{\mu\nu} \hat R^{\mu\nu} + c_6 \hat R_{\mu}^0 \hat R^{\mu0} \right) \,.
\end{equation}
(Here the index 0 means contracting with $\partial \chi, \hat{R}^0_\mu \equiv \hat{R}^\lambda_{~\mu} \partial_\lambda \chi$.) The procedure follows closely the one for the current and we will only highlight the main points. 

The stress-energy tensor can be computed considering linear perturbations around flat space, $T^{\mu \nu} = (-g)^{-1/2} \delta S / \delta g_{\mu \nu} \big|_{g = \eta}$ (we discard the conventional factor of 2). We are interested in the correlator
\begin{equation}
	\langle T^{\mu \nu}(-k) T^{\rho \sigma}(k) \rangle  \,,
\end{equation}
which, as in the $\langle JJ \rangle$ computation of \S\ref{JJsec}, will involve contact terms generated by the $\mathcal{O}(\delta g^2)$ terms in the action. (Constant terms in $T^{\mu \nu}$ drop out when taking the commutator in the Green's function.)

It is useful to discuss the conservation and symmetry properties of $\langle T^{\mu \nu}(-k) T^{\rho \sigma}(k) \rangle$. The most general conserved and traceless object $\langle T^{\mu \nu} T^{\rho \sigma} \rangle_\textsf{subl.}$ (the subscript ${}_\textsf{subl.}$ means that we focus on the NLO and NNLO parts) with the correct symmetries ($\mu \leftrightarrow \nu$, $\rho \leftrightarrow \sigma$, $(\mu \nu) \leftrightarrow (\rho \sigma)$) in $d > 2$ dimensions can be written as a linear combination of two structures,
\begin{equation} \label{Tmunugeneral}
	i \langle T^{\mu \nu} (-k) T^{\rho \sigma} (k) \rangle_\textsf{subl.} = \textsf{C}(k) \, \Pi^{\mu \nu \rho \sigma}(k) + \textsf{D}(k) \, \tilde{\Pi}^{\mu \nu \rho \sigma} (k)
\end{equation}
with
\begin{align}
	\Pi^{\mu \nu \rho \sigma} &= \frac{1}{2} \left( \pi^{\mu \rho} \pi^{\nu \sigma} + \pi^{\mu \sigma} \pi^{\nu \rho} \right) - \frac{1}{d-1} \pi^{\mu \nu} \pi^{\rho \sigma} \,, \\
	\tilde{\Pi}^{\mu \nu \rho \sigma} &= \frac{1}{4} \left( \pi^{\mu \rho} \tilde{\pi}^{\nu \sigma} + \pi^{\mu \sigma} \tilde{\pi}^{\nu \rho} +\pi^{\nu \sigma} \tilde{\pi}^{\mu \rho} + \pi^{\nu \rho} \tilde{\pi}^{\mu \sigma}  \right) - \frac{1}{d-2} \tilde{\pi}^{\mu \nu} \tilde{\pi}^{\rho \sigma} \,,
\end{align}
where
\begin{align}
	\pi^{\mu \nu} &\equiv \eta^{\mu \nu} - \frac{k^\mu k^\nu}{k^2} \,, \\
	\tilde{\pi}^{\mu \nu} &= \delta^{mn} - \frac{k^m k^n}{\boldsymbol{k}^2} \,.
\end{align}
The required symmetries and conservation are manifest in these formulas.\footnote{To show tracelessness the following identities are useful:
\begin{align}
	\pi &= \pi_\mu^{~\mu} = d-1 \,, \tilde{\pi} = d-2 \,, \\
	\eta^{\mu \nu} \pi^{\mu \rho} \pi^{\nu \sigma} &= \pi^{\rho \sigma} \,, \\
	\eta^{\mu \nu} \pi^{\mu \rho} \tilde{\pi}^{\nu \sigma} &= \tilde{\pi}^{\rho \sigma} \,.
\end{align}}
The operator proportional to $c_1$ gives a contribution with a different structure as a consequence of the vacuum expectation of $\langle T_{\mu\nu}\rangle$: this is the reason for the subscript $\textsf{subl.}$, since the leading term must be dropped. We are not considering the operator $c_1$ for the positivity discussion, so we postpone the discussion of this ``modified'' conservation to Appendix \ref{app:conservation}.

It is a straightforward, but very long, exercise to get the functions $\textsf{C}$ and $\textsf{D}$ for the action of Eq.~\eqref{EFThataction}:
\begin{align}
	\textsf{C} &= -\frac{\mu}{2} \frac{\omega^2 (\omega^2 - \boldsymbol{k}^2)^2}{(\omega^2 - c_s^2 \boldsymbol{k}^2)^2} (c_2 + c_3) + \frac{1}{\mu} \frac{\boldsymbol{k}^4 (\omega^2 - \boldsymbol{k}^2)^2}{(\omega^2 - c_s^2 \boldsymbol{k}^2)^2} c_4 + \frac{1}{2\mu} \frac{(\omega^2 - \boldsymbol{k}^2)^2 \left( \omega^2 (\omega^2 - \boldsymbol{k}^2) + \boldsymbol{k}^4 \right)}{(\omega^2 - c_s^2 \boldsymbol{k}^2)^2} c_5 \notag \\ &+ \frac{1}{4\mu} \frac{\boldsymbol{k}^2 \omega^2 (\omega^2 - \boldsymbol{k}^2)^2}{(\omega^2 - c_s^2 \boldsymbol{k}^2)^2} c_6 - \frac{1}{2\mu} \frac{(c_2 + c_3)^2}{c_1} \frac{\boldsymbol{k}^4 \omega^2 (\omega^2 - \boldsymbol{k}^2)^2}{(\omega^2 - c_s^2 \boldsymbol{k}^2)^3} \,, \label{ATT} \\
	\textsf{D} &= -\frac{\mu}{4} \frac{\boldsymbol{k}^4 (\omega^2 - \boldsymbol{k}^2)}{(\omega^2 - c_s^2 \boldsymbol{k}^2)^2} (c_2 + c_3) - \frac{1}{\mu} \frac{\boldsymbol{k}^4 (\omega^2 - \boldsymbol{k}^2)^2}{(\omega^2 - c_s^2 \boldsymbol{k}^2)^2} \left( 2c_4 + \frac{3}{4} c_5 \right) + \frac{1}{8\mu} \frac{\boldsymbol{k}^6 (\omega^2 - \boldsymbol{k}^2)}{(\omega^2 - c_s^2 \boldsymbol{k}^2)^2} c_6 \notag \\
	&+\frac{1}{\mu} \frac{(c_2 + c_3)^2}{c_1} \frac{\boldsymbol{k}^4 \omega^2 (\omega^2 - \boldsymbol{k}^2)^2}{(\omega^2 - c_s^2 \boldsymbol{k}^2)^3} \,. \label{BTT}
\end{align}

\subsection{Positivity bounds from $\langle TT \rangle$ \label{TTpossec}}
Following the $\langle JJ \rangle$ computation of \S\ref{JJpossec}, we contract~\eqref{Tmunugeneral} with two copies of a general symmetric two-tensor $A$: $\langle T^{\mu \nu} T^{\rho \sigma} \rangle A_{\mu \nu} A_{\rho \sigma}$. We take $\boldsymbol{k}_0 = \boldsymbol{0}$  and expand $A$ with constant coefficients as
\begin{equation}
	A_{\mu \nu} = \alpha \hat{K}_\mu \hat{K}_\nu + \beta \hat{E}_\mu \hat{E}_\nu + \gamma \hat{F}_\mu \hat{F}_\nu + \tilde{\alpha} \left( \hat{K}_\mu \hat{E}_\nu + \hat{K}_\nu \hat{E}_\mu \right) + \tilde{\beta} \left( \hat{K}_\mu \hat{F}_\nu + \hat{K}_\nu \hat{F}_\mu \right) + \tilde{\gamma} \left( \hat{E}_\mu \hat{F}_\nu + \hat{E}_\nu \hat{F}_\mu \right) \,,
\end{equation}
where $\hat{K}, \hat{E}$ and $\hat{F}$ are given in Eqns.~\eqref{Khat}-\eqref{Fhat}. The following identities are useful in the calculation of the contraction: $\pi^{\mu \alpha} \hat{K}_\mu = \tilde{\pi}^{\mu \alpha} \hat{K}_\mu = \tilde{\pi}^{\mu \alpha} \hat{E}_\mu = 0, \pi^{\mu \alpha} \hat{E}_\mu = \hat{E}^\alpha \,, \pi^{\mu \alpha} \hat{F}_\mu = \hat{F}^\alpha$ and $\tilde{\pi}^{\mu \alpha} \hat{F}_\mu = \hat{F}^a$. A straightforward calculation then yields
\begin{equation} \label{TTAA}
	i \langle T^{\mu \nu} T^{\rho \sigma} \rangle_\textsf{subl.} A_{\mu \nu} A_{\rho \sigma} = \frac{\textsf{C}}{2} \left[ (\beta - \gamma)^2 + 4 \tilde{\gamma}^2 \right] + \textsf{D} \tilde{\gamma}^2 \,.
\end{equation}

In the CFT the two-point function of $T^{\mu\nu}$ in Fourier space goes as $\omega^d$; therefore we must divide by at least $\omega^5$ in $d = 3$ to be able to neglect the contour at infinity. Setting $\boldsymbol{k} = \omega \boldsymbol{\xi}$, dividing by $\omega^5$ and integrating around a small circular contour going around the origin in the counterclockwise direction, we obtain the positivity conditions. Defining $\delta \equiv \beta - \gamma$ since only this combination involving $\beta$ and $\gamma$ appears in~\eqref{TTAA}, we find
\begin{align}
	4 \xi^4 \delta^2 c_4 +
	2 \left[(2-\xi^2)^2 \tilde{\gamma}^2 + (1-\xi^2+\xi^4) \delta^2 \right] c_5 &+ \xi^2 \left( \frac{(2-\xi^2)^2}{1-\xi^2} \tilde{\gamma}^2 + \delta^2 \right) c_6 \notag \\
	&\geq \frac{4 \xi^4 \delta^2}{2-\xi^2} \frac{(c_2 + c_3)^2}{c_1} \label{ineq125}
\end{align}
for all $\delta, \tilde{\gamma}$ and $\xi \in [0,1)$. Again this collection of bounds may be reformulated in a more insightful way. By choosing $\delta = 0$, we obtain $2(1-\xi^2) c_5 + \xi^2 c_6 \geq 0$. Taking $\xi = 0$ and $\xi = 1$ here gives $\boxed{c_5 \geq 0}$ and $\boxed{c_6 \geq 0}$, which are equivalent to the bounds for all values of $\xi$. Since $c_5$ and $c_6$ are positive, so that $\tilde{\gamma}^2$ appears on the LHS multiplying something positive, the strongest bound is obtained setting $\tilde{\gamma} = 0$. This gives the bounds
\begin{equation} \label{c457constraints}
	4 \xi^4 c_4 + 2(1-\xi^2+\xi^4) c_5 + \xi^2 c_6 \geq \frac{4 \xi^4}{2-\xi^2} \frac{(c_2 + c_3)^2}{c_1} \,.
\end{equation}
It is not hard to show that the most stringent of these bounds is attained when $\xi = 1$: $\boxed{4 c_4+2c_5+c_6\geq 4 (c_2+c_3)^2/c_1}$. \\

One could also derive additional bounds by studying the positivity of linear combinations of $T^{\mu\nu}$ and $J^\mu$. Since the two operators have different dimensions, one could consider a linear combination of the form $\mu J + T$, where $\mu$ is the chemical potential. (We are here very schematic and drop indices: of course one should also introduce a vector to match indices and vary over this vector to get the most general constraint.) In this case, for the convergence of the stress-energy tensor part, one needs to go to NNLO, so that we would need to consider also the NNLO contact operators for the current (the analogue of the operators proportional to $b$ and $d$ but with two more derivatives).  Another possibility would be to consider combinations of the schematic form $i \omega J + T$: in this case, by dividing the two-point function by $\omega^5$ one would select the NLO in the $\langle JJ \rangle $ part and NNLO in $\langle TT \rangle$, i.e.~the operators we have been considering so far. However nothing new comes from the cross-terms $\langle JT \rangle$, at least for the operators that we studied. Indeed, it is straightforward to realize that all our two-point functions, including the $\langle JT\rangle$ one, are even in $\omega$, so that one does not get any simple pole after the multiplication by $i \omega$. The reason of this parity in $\omega$ can be explained in terms of symmetry. The $\chi$ action is time-reversal symmetric, but the background $\chi = \mu t$ spontaneously breaks this symmetry. If, however, the action is also invariant under $\chi \to - \chi$, then there is a residual unbroken symmetry: $t \to - t$, $\pi \to - \pi$. This symmetry imposes to have only an even number of $\omega$'s in a two-point function. We did not impose by hand any $\chi \to -\chi$ symmetry, but at the level of the operators we considered it turns out to be an accidental symmetry: $\hat g$ contains an even number of $\chi$'s and we always considered operators with an even number of extra $\chi$'s. Looking at operators with even more derivatives, one finds terms that break this symmetry: for instance $\hat R \hat R^{\mu\nu} \hat\nabla_\nu\hat\nabla_\mu\chi$ contains an odd number of $\chi$'s. Even at the order we are working there is a contact operator for $\langle JT \rangle$ of the form $\hat R^{\mu 0} F^0_{~\mu}$ that contributes to the two-point function $\langle JT \rangle $ with an odd number of $\omega$'s. The coefficient of this operator would be constrained by the mixed correlator and the same will happen in a general theory without $\chi \to - \chi$ symmetry. We leave this study to future work.

\subsection{Summary of bounds \label{summarybounds}}
In summary, the bounds we have been able to obtain on the NLO and NNLO coefficients of the EFT of conformal superfluids are
\begin{align}
	c_1 &\geq 0 \,\quad {\rm (for\;healthy\;fluctuations)}, \label{c1ineq} \\
	\frac{c_2}{b+d} (1-\xi^2) - \frac{c_3}{b+d} &\geq -\frac{(1-\xi^2/2)^2}{\xi^2} \,, \label{c2c3constraints} \\
	d &\geq 0 \,, \\
	b+d &\geq 0 \,, \\
	4 c_4+2c_5+c_6 &\geq 4 (c_2+c_3)^2/c_1 \,, \label{4c4+2c5} \\
	c_5 &\geq 0 \,, \\
	c_6 &\geq 0 \,. \label{c6ineq}
\end{align}
The constraints in~\eqref{c2c3constraints} hold for all $\xi \in [0,1)$ and are plotted in Fig.~\ref{c2c3fig}. The action of this EFT for the Goldstone boson $\pi$ corresponding to the spontaneous breaking of Lorentz boosts is given by Eq.~\eqref{EFThataction} where $\hat{g}_{\mu \nu} = g_{\mu \nu} |g^{\alpha \beta} \partial_\alpha \chi \partial_\beta \chi|$ and $\chi = \mu t + \pi$, the first three terms of which we have written out explicitly in Eq.~\eqref{gaugeL}.

\subsection{Loop corrections?\label{sec:loops}}
Our calculations so far have been at tree level in the EFT, but there are no obstacles to including loops in our formalism. In general, loops, in the absence of a mass gap, will open a cut in the $\omega$-plane all along the real axis. As discussed above, the contours of Fig.~\ref{fig:doublecontour} are applicable in this case. In the EFT one can calculate the integral over the two small semicircles: the radius of these semicircles does not need to be infinitesimal, it is enough that it is small enough for the EFT calculation to be reliable. In general, the result will depend on the radius and this dependence reproduces the scale dependence induced by loops. The contributions of these arcs are related via Cauchy's theorem to the remaining integrals along the real axis, which are constrained to be positive. The general picture is quite similar to what happens in the Lorentz-invariant case, see \cite{Bellazzini:2020cot}.

It turns out, however, that for the particular example we are studying loops are actually absent, so that the bounds derived above are sharp and not approximate. To understand why, let us rewrite schematically the action for $\pi$ in canonical normalization, including the interaction terms,
\begin{align}
	\mathcal{L} &=  \frac{\mu c_1}{2} \left[ \dot{\pi}^2 - \frac{1}{2} (\partial_i \pi)^2  \right] + c_1 \dot\pi^3 + \frac{c_1}{\mu} \dot\pi^4 + \ldots \frac{c_{2;3}}{\mu} \left[ \partial^2 \pi \partial^2 \pi +  \partial^2 \pi \partial^2 \pi \dot\pi + \ldots \right] + \notag  \\ & + \frac{c_{4;5;6}}{\mu^3} \left[ \partial^3 \pi \partial^3 \pi +   \ldots \right] 
	&\notag \\ &=  \frac{1}{2} \left[ \dot{\pi_c}^2 - \frac{1}{2} (\partial_i \pi_c)^2  \right] + \frac{1}{c_1^{1/2} \mu^{3/2}}\dot\pi_c^3 + \frac{1}{c_1 \mu^3} \dot\pi_c^4  + \frac{c_{2;3}}{c_1 \mu^2} \partial^2 \pi_c \partial^2 \pi_c +  \frac{c_{2;3}}{c_1^{3/2} \mu^3} \partial^2 \pi_c \partial^2 \pi_c \dot\pi_c &\notag \\ & + \frac{c_{4;5;6}}{c_1 \mu^4} \partial^3 \pi_c \partial^3 \pi_c + \ldots
\end{align}
From this expression in canonical normalization, it is clear that one cannot get loop corrections with same $\mu$-dependence as the one induced by $c_{2;3}$ and $c_{4;5;6}$ (in particular this says these coefficients do not run). Indeed, looking at the powers of $\mu$, one sees it is impossible for the interactions, both the ones coming from the $c_1$ operator and the ones from the $c_{2;3}$ terms, to combine to give the quadratic terms proportional to $c_{2;3} /\mu^{2}$  and $c_{4;5;6}/ \mu^{4}$. (Notice that combining two $\dot\pi^4$ interactions one gets $\mu^{-6}$ which corresponds to operators with two more derivatives with respect to  $c_{4;5;6}$: at this order one starts having log-divergences.)

\section{UV complete example: conformal scalar in $d = 3$} \label{3DUVsec}
In this section we test our constraints~\eqref{c1ineq}-\eqref{c6ineq} in a simple explicit UV complete example, a conformal complex scalar in $d = 3$. The UV action is 
\begin{align}
	\mathcal{L}_\textsf{UV} &= \sqrt{-g} \left( -|\partial \phi|^2 - \lambda |\phi|^6 - \frac{1}{8} R |\phi|^2 \right) \notag \\
	&= \sqrt{-g} \left( -(\partial \rho)^2 + \rho^2 |\partial \theta|^2 - \lambda \rho^6 - \frac{1}{8} R \rho^2 \right) \,, \label{UVLagrangian}
\end{align}
where $\phi = \rho \, e^{i \theta}$ and $\lambda > 0$ (a more general model is studied in Appendix \ref{UV2scalars}). This theory is conformally invariant at tree level and it has been studied in detail in \cite{Badel:2019khk}. We will be interested in a state with finite chemical potential, i.e.~with $\theta = \mu t + \pi(t,\boldsymbol{x})$ and work at leading order in $\lambda$ so that the theory is conformal. To derive the EFT action at tree level we integrate out the radial field $\rho$ (we require $\lambda \neq 0$ so that $\rho$ has a finite mass). Its equation of motion is
\begin{equation}
	\square \rho + |\partial \theta|^2 \rho - 3\lambda \rho^5 - \frac{R}{8} \rho = 0 \,.
\end{equation}
In a derivative expansion we have the solution $\rho = \rho_0 + \rho_1 + \rho_2 + \cdots$, where
\begin{equation}
	\rho_0 = \frac{|\partial \theta|^{1/2}}{(3\lambda)^{1/4}} \,, ~~ \rho_1 = \frac{\square \rho_0 - R \rho_0/8}{4 |\partial \theta|^2} \,,
\end{equation}
where we treated $R$ as a two-derivative term. To obtain the EFT action up to terms involving two more $\square$'s than the leading term, we insert $\rho$ into the action (it is not hard to see that $\rho_2$ is not required for this). Some manipulations then yield
\begin{equation}
	\frac{\mathcal{L}_\textsf{EFT}}{\sqrt{-g}} = \frac{2}{3 \sqrt{3\lambda}} |\partial \theta|^3 - \frac{1}{8 \sqrt{3\lambda}} \left( |\partial \theta| R + 2 \frac{(\partial |\partial \theta|)^2}{|\partial \theta|} \right) + \frac{1}{4 |\partial \theta|^2} \left( \square \rho_0 - \frac{R}{8} \rho_0 \right)^2 \,.
\end{equation}
From here we see
\begin{equation} \label{UVcoefficients}
	c_1 = \frac{4}{\sqrt{3 \lambda}} \,, ~~ c_2 = \frac{1}{8 \sqrt{3 \lambda}} \,, ~~ c_4 = \frac{1}{256 \sqrt{3 \lambda}} \,, ~~ c_ 3 = c_5 = c_6 = b = d = 0 \,.
\end{equation}
It may be checked that all bounds~\eqref{c1ineq}-\eqref{c6ineq} are satisfied by these choices of coefficients, in particular all bounds are saturated except for~\eqref{c1ineq} and~\eqref{c2c3constraints}.

\paragraph{Calculation of UV correlators.} In the theory~\eqref{UVLagrangian} we can calculate the correlators $\langle JJ \rangle$ and $\langle TT \rangle$ at tree level, verifying that they indeed reduce to the general EFT expressions~\eqref{JmuJnu} and~\eqref{Tmunugeneral} when expanded to NLO and NNLO in $\omega/\mu$ respectively, for the choice of coefficients~\eqref{UVcoefficients}.

For $\langle JJ \rangle$ we have the conserved current
\begin{equation}
	J_\mu = i \left( \phi \, \partial_\mu \phi^* - \phi^* \partial_\mu \phi \right) = 2 \rho^2 \partial_\mu \theta
\end{equation}
associated to the $U(1)$ symmetry $\phi \rightarrow e^{i \alpha} \phi$. To calculate $\langle JJ \rangle$ we couple $\phi$ to an external gauge field $A_\mu$, replacing $\partial_\mu \rightarrow \nabla_\mu$. This introduces extra terms $J_\mu A^\mu - \rho^2 A^2$ in the action~\eqref{UVLagrangian}. Then we set 
\begin{align}
	\theta &= \mu t + \frac{\pi}{\sqrt{2} \hat{\rho}} \,, \label{mutpluspi} \\
	\rho &= \hat{\rho} + \frac{\delta \rho}{\sqrt{2}} \,, \label{rhohatplusdeltarho}
\end{align}
where $\hat{\rho} = |\partial \theta|^{1/2} / (3\lambda)^{1/4} \approx \sqrt{\mu} / (3\lambda)^{1/4}$, obtaining the canonically normalized quadratic action
\begin{align}
	\mathcal{L}_{(2)}(x) &= - \frac{1}{2} \left( \partial \pi \right)^2 - \frac{1}{2} \left( \partial \delta \rho \right)^2 - 2 \mu^2 \delta \rho^2 + 2 \mu \, \delta \rho \, \dot{\pi} \,.
\end{align}
In Fourier space this reads
\begin{align}
	\mathcal{L}_{(2)}(k) &= \frac{1}{2} \begin{pmatrix} \tilde{\pi}(-k) & \tilde{\delta \rho}(-k) \end{pmatrix} \begin{pmatrix} -k^2 & 2 i \mu \omega \\ -2 i \mu \omega  & -k^2 - 4 \mu^2 \end{pmatrix} \begin{pmatrix} \tilde{\pi}(k) \\ \tilde{\delta \rho}(k) \end{pmatrix} \,. \label{pideltarhoL}
\end{align}
Now we have, to first order in the fields,
\begin{align}
	J^0 &= - \frac{2 \mu^2}{\sqrt{3 \lambda}} - \sqrt{2} \hat{\rho} \left( \dot{\pi} + 2 \mu \, \delta\rho \right) \,, \label{J0UV1st} \\
	J^i &= \sqrt{2} \hat{\rho} \, \partial_i \pi \,. \label{JiUV1st}
\end{align}
With this we find
\begin{align}
	i \langle J^\mu(-k) J^\nu(k)\rangle = \textsf{A}_{\text{UV}} \left( k^\mu k^\nu - \eta^{\mu \nu} k^2 \right) + \textsf{B}_{\text{UV}} \left( k^i k^j - \delta^{ij} \boldsymbol{k}^2 \right) \,,\label{JmuJnuUV}
\end{align}
where
\begin{align}
	\textsf{A}_{\text{UV}} &= -\frac{2 \mu}{\sqrt{3 \lambda}} \frac{\omega^2 - \boldsymbol{k}^2 - 8\mu^2}{(\omega^2 - \boldsymbol{k}^2)(\omega^2 - \boldsymbol{k}^2 - 4\mu^2) -4\mu^2 \omega^2} \,, \\
	\textsf{B}_{\text{UV}} &= -\frac{2 \mu^3}{\sqrt{3 \lambda}} \frac{4}{(\omega^2 - \boldsymbol{k}^2)(\omega^2 - \boldsymbol{k}^2 - 4\mu^2) - 4\mu^2 \omega^2} \,.
\end{align}
Expanding $\textsf{A}_{\text{UV}}$ and $\textsf{B}_{\text{UV}}$ to first subleading order in $\omega/\mu$ and making the choice of coefficients~\eqref{UVcoefficients}, we observe that $\textsf{A}_{\text{UV}}$ and $\textsf{B}_{\text{UV}}$ reduce to their EFT counterparts $\textsf{A}$ and $\textsf{B}$ in~\eqref{JJAEFT}-\eqref{JJBEFT} indeed. In this particular example the UV contribution to the contour argument of Fig.~\ref{fig:contour} is simply given by the two UV poles associated with the massive radial mode.\footnote{At the order $1/\sqrt{\lambda}$ at which we are working, the correlator never asymptotes to the unbroken CFT result~\eqref{eq:JJCFT}, because we can never neglect the vev of $\rho$ at this order. $\langle JJ\rangle$ clearly does go to the CFT result in the UV at order $\lambda^0$. Therefore, for the contour argument at order $1/\sqrt{\lambda}$, we can use the expressions~\eqref{JmuJnuUV}.}

We now perform the same exercise for $\langle TT \rangle$. The theory~\eqref{UVLagrangian} is already exhibits a conformal coupling to $g$, and we compute $T_{\mu \nu}$ by expanding to linear order around flat space (again, as in the EFT computation, we discard a conventional factor of 2),
\begin{equation}
	T_{\mu \nu} = \frac{1}{2} 
	\left( \partial_\mu \phi^* \, \partial_\nu \phi + \partial_\mu \phi \, \partial_\nu \phi^* \right) + \frac{1}{2} \eta_{\mu \nu} \mathcal{L}_\textsf{UV} + \frac{1}{8} \left( \eta_{\mu \nu} \Box |\phi|^2 - \partial_\mu \partial_\nu |\phi|^2 \right) \,.
\end{equation}
As before, quadratic terms in $\delta g$ will be important for the correct computation of $\langle TT \rangle$. As discussed in Appendix \ref{app:conservation}, the entire UV correlator $\langle T^{\mu \nu}(-k) T^{\rho \sigma}(k) \rangle$ contains a piece which is not conserved in the  ``usual'' way as a consequence of the presence of a background $\langle T^{\mu \nu} \rangle$. To get a transverse structure we must subtract off the leading term in $\omega/\mu$, which corresponds to the term proportional to the coefficient $c_1$ in the EFT. After doing this one gets
\begin{equation} \label{TmunugeneralUV}
	i \langle T^{\mu \nu} (-k) T^{\rho \sigma} (k) \rangle_\textsf{subl.} = \textsf{C}_\text{UV}(k) \, \Pi^{\mu \nu \rho \sigma}(k) + \textsf{D}_\text{UV}(k) \, \tilde{\Pi}^{\mu \nu \rho \sigma} (k) \,,
\end{equation}
like in the EFT result~\eqref{Tmunugeneral}, but with
\begin{align}\label{AUVTT}
	\textsf{C}_{\text{UV}} &= -\frac{\mu}{32 \sqrt{3 \lambda}} \frac{(\omega^2 - \boldsymbol{k}^2)^2 \left[ (\omega^2 - \boldsymbol{k}^2)^2 + \omega^2 \left( \omega^2 - \boldsymbol{k}^2 - 16 m^2 \right) \right]}{\left( \omega^2 - c_s^2 \boldsymbol{k}^2 \right) \left[ (\omega^2 - \boldsymbol{k}^2)(\omega^2 - \boldsymbol{k}^2 - 4\mu^2) - 4\mu^2 \omega^2 \right]} \,, \\ \label{BUVTT}
	\textsf{D}_{\text{UV}} &= -\frac{\mu^3}{4 \sqrt{3 \lambda}} \frac{\boldsymbol{k}^4 (\omega^2 - \boldsymbol{k}^2)}{\left( \omega^2 - c_s^2 \boldsymbol{k}^2 \right) \left[ (\omega^2 - \boldsymbol{k}^2)(\omega^2 - \boldsymbol{k}^2 - 4\mu^2) - 4\mu^2 \omega^2 \right]} \,.
\end{align}
Again, the correspondence between $\textsf{C}_{\text{UV}},\textsf{D}_{\text{UV}}$ and their EFT counterparts $\textsf{C}, \textsf{D}$ in~\eqref{ATT}-\eqref{BTT} may be verified, this time to NNLO in $\omega/\mu$. Again, the UV contribution to the imaginary part is just due to the poles of the massive radial direction.

\paragraph{Note on the saturation of bounds.} It is at first puzzling that, except for~\eqref{c1ineq} and~\eqref{c2c3constraints}, all of our general positivity inequalities~\eqref{c1ineq}-\eqref{c6ineq} are saturated for this particular UV theory. To understand why this is the case, consider the derivation of e.g.~\eqref{4c4+2c5}, assuming $c_3 = c_5 = c_6 = 0$. This inequality arises from considering the function $\langle T^{\mu \nu} T^{\rho \sigma} \rangle A_{\mu \nu} A_{\rho \sigma}$ on the LHS of~\eqref{TTAA}, setting $\boldsymbol{k} = \omega \boldsymbol{\xi}$, dividing by $\omega^5$ and integrating counterclockwise around a small circle $\mathcal{C}_0$ centered at $\omega = 0$. That is
\begin{equation} \label{IC0}
	\mathcal{I} = \int_{\mathcal{C}_0} \di \omega \frac{i \langle T^{\mu \nu} T^{\rho \sigma} \rangle A_{\mu \nu} A_{\rho \sigma}}{\omega^5} \,.
\end{equation}
In a general EFT with $c_3 = c_5 = c_6 = 0$, this integral may be evaluated using the residue theorem for the pole at $\omega = 0$, from which we obtain (cf.~\eqref{ineq125})
\begin{equation} \label{IEFT}
	\mathcal{I} = \frac{4 \pi i}{\mu} \frac{\xi^4 (1-\xi^2)^2}{(2-\xi^2)^2} \left( c_4 - \frac{1}{2-\xi^2}\frac{c_2^2}{c_1} \right) \delta^2 \,.
\end{equation}
Here we also know the integrand in~\eqref{IC0} at large $\omega$, and can explicitly calculate the UV contributions, from the UV poles located at the zeroes of the denominator in Eq.~\eqref{AUVTT},
\begin{equation}
	\omega_\pm = \pm \frac{2 \mu \sqrt{2-\xi^2}}{1-\xi^2} \,,
\end{equation}
using Cauchy's theorem. This way we obtain
\begin{equation} \label{IUV}
	\mathcal{I} = \frac{i \pi}{64 \mu \sqrt{3 \lambda}} \frac{\xi^4 (1-\xi^2)^3}{(2-\xi^2)^3} \delta^2 \,,
\end{equation}
which is indeed of the form $i \times \text{(positive)}$, as it should be from our general discussion in \S\ref{sec:setup}. Moreover one may check that filling in the appropriate values for $c_{1,2,4}$ in~\eqref{IEFT} yields the result~\eqref{IUV} indeed. Equating~\eqref{IEFT} with~\eqref{IUV} gives
\begin{equation}
	c_4 - \frac{1}{2-\xi^2} \frac{c_2^2}{c_1} = \frac{1}{16^2 \sqrt{3\lambda}} \frac{1-\xi^2}{2-\xi^2} \,.
\end{equation}
Now, without specific knowledge of the UV contribution, we would have deduced that
\begin{equation}
	c_4 = \frac{1}{2-\xi^2} \frac{c_2^2}{c_1} + (\text{UV}) \geq \frac{1}{2-\xi^2} \frac{c_2^2}{c_1} \,,
\end{equation}
as we did in~\eqref{ineq125}. The RHS of this inequality becomes largest, so the bound is most stringent, when we send $\xi \rightarrow 1$. But in this limit $(\text{UV}) \rightarrow 0$. This is why the inequality is saturated in this case.

In Appendix \ref{UV2scalars} we consider a two-field UV theory where the bound~\eqref{4c4+2c5} is not saturated.

\section{Conclusions and future directions\label{sec:conclusions}}
In this paper we presented a method to derive positivity constraints on EFT's with spontaneously broken Lorentz invariance. It is based on the study of the two-point function of a conserved current or the stress-energy tensor. The analytic properties of these objects, together with the assumption that the theory becomes conformal in the deep UV, allow to run a dispersive argument similar to the one for the $S$-matrix in the case of Lorentz invariant theories. We applied this method to the particular example of conformal superfluids in three dimensions and we checked that the inequalities we derived on the coefficients of the EFT are indeed satisfied in simple perturbative tree-level UV completions.

It would be reassuring to verify our inequalities when the UV completion is more complicated. For instance one could go beyond tree level in the $\lambda |\phi|^6$ theory discussed above: indeed the theory remains conformal at one loop \cite{Badel:2019khk}, so one could verify our contour arguments including the contribution of the high-energy cuts. The EFT of conformal superfluids applies also when the UV completion is a strongly coupled CFT. In this case one could derive analytically the coefficients of the EFT in a $1/N$ expansion. This has been done for some CFT data, like the dimension of the lowest operators of given charge \cite{Giombi:2020enj}, and the same should be possible for all the coefficients that appear in the EFT. 

Leaving aside conformal superfluids one can apply our arguments to realistic systems. One possibility is the color-flavor-locking (CFL) phase of QCD at high chemical potential \cite{Alford:1998mk}. This phase is quite similar to a conformal superfluid, but the underlying theory is clearly not conformal since it contains the additional scale $\Lambda_{\rm QCD}$. Our method is applicable anyway since we only need to control the UV behavior of the Green's functions. At a more basic level one can go back to the venerable problem of the propagation of light in a material: to our knowledge the full set of constraints deriving from the relativistic version of the Kramers-Kronig relation \cite{MS76} together with the requirement that the medium can only absorb energy from the wave have not been explicitly spelled out. It should also be possible to use our general arguments to constrain the coefficients that enter in the description of dissipative fluid dynamics (see for instance \cite{Bhattacharya:2011tra}). 

Our results are robust but the constraints we derived are rather limited. Since we are using two-point functions as fundamental objects, we are sensitive to operators that contribute to the quadratic action (in general these operators are forced by symmetry to contain non-linear terms as well): we cannot say anything about operators that start cubic, or higher, in fluctuations. Moreover one needs to go sufficiently high in derivatives, in such a way that the contour integral at infinity, where the theory can be approximated by a CFT, converges. It would be nice to extend our bounds to operators with more perturbations and/or less derivatives. One natural extension is to generalize our arguments to higher-order correlation functions, say $\braket{JJJJ}$. It should also be possible to relax the assumption of CFT in the UV: for instance it should be enough that the theory is perturbative at scales well above the scale of Lorentz breaking. Work is in progress in all these directions.

We just started to uncover the full set of constraints on theories with spontaneous breaking of Lorentz invariance: clearly much work remains to be done.

\paragraph*{Acknowledgements.}
It is a pleasure to thank B.~Bellazzini, G.~Cuomo, L.~Delacretaz, J.~Elias-Mir\'o, A.~Longo, S.~Melville, S.~Minwalla, M.~Mirbabayi, A.~Nicolis, R.~Rattazzi, P.~Romatschke, L.~Santoni, S.~Shenker, E.~Silverstein, M.~Serone and G.~Villadoro for useful discussions.

\appendix

\section{EFT operators} \label{EFToper}
In this appendix we aim to obtain the independent operators of the EFT  at each order in the derivative expansion. The simplest way to write operators that non-linearly realize the full conformal group is to use the modified metric $\hat g_{\mu\nu} \equiv g_{\mu\nu}| g^{\alpha\beta}\partial_\alpha\chi\partial_\beta\chi |$ \cite{Cuomo:2020rgt}, where $\chi = \mu t +\pi(t,x)$. The metric $\hat g_{\mu\nu}$ is Weyl invariant so that any diffeomorphism invariant action built using this metric, instead of $g_{\mu\nu}$, will automatically non-linearly realize the whole conformal group.\footnote{In fact, a conformal transformation combined with a suitable diffeomorphism amounts to a Weyl rescaling of the metric. The conformal group is non-linearly realized because $\chi$ has a vev.} 
At leading order in derivatives we have only the term
\begin{equation}\label{eq:S1}
S^{(1)} = \frac{c_1}{6} \int \di^3 x \sqrt{-\hat g} = \frac{c_1}{6} \int \di^3 x \sqrt{-g} |\partial\chi|^3 \,.
\end{equation}
At next order we have to add two derivatives to the action.\footnote{We assume parity invariance, otherwise one could write a term with a single extra derivative \cite{Cuomo:2021qws}.} One can write two operators using the Ricci tensor:
\begin{equation}\label{eq:S2}
S^{(2)} =  \int \di^3 x \sqrt{-\hat g} \left(-c_2 \hat R + c_3 \hat R^{\mu\nu} \hat\partial_\mu \chi \hat\partial_\nu\chi \right) \,.
\end{equation}
(There is no difference between $\partial_\mu$ and $\hat\partial_\mu$, but we use this notation to emphasize that indices are raised and contracted with $\hat g_{\mu\nu}$.) These two are the only operators at this order, as we are now going to show. The only other operator one has to consider is $\int \di^3 x \sqrt{-\hat g} \hat \Box \chi \hat \Box \chi$. (One can get rid of terms with three derivatives acting on a single $\chi$ integrating by parts, so that one can focus on operators of the schematic form $(\hat\nabla\hat\nabla\chi)^2$, besides the terms with Ricci written above. One cannot contract indices with $\hat\partial\chi$ since $\hat\nabla_\mu\hat\nabla^\alpha\chi \hat\partial_\alpha\chi = \frac12 \hat\partial_\mu (\hat\partial^\alpha\chi \hat\partial_\alpha\chi)=0$, so one is left with $\int \di^3 x \sqrt{-\hat g} \hat \Box \chi \hat \Box \chi$ and $\int \di^3 x \sqrt{-\hat g} \hat \nabla_\mu \hat \nabla_\nu \chi \hat \nabla^\mu \hat \nabla^\nu \chi$, which are equivalent integrating by parts.) One can dispose of this additional operator by a field redefinition. Indeed consider a perturbative field redefinition
\begin{equation}\label{eq:fieldred}
\tilde\chi = \chi + \varepsilon\;\hat\Box\chi\,.
\end{equation}
Since $\hat\Box\chi$ is Weyl invariant $\chi$ and $\tilde\chi$ have the same transformation properties: one could use $\tilde\chi$ instead of $\chi$ to build the metric $\hat g_{\mu\nu}$ and construct out of it the operators in the action. One can thus use this field redefinition in the leading action Eq.~\eqref{eq:S1} to get rid of an operator. The field redefinition will generate a term proportional to the equation of motion:
\bea\nn
&&\int \di^3 x \sqrt{-g} |\partial\chi|^3 \to -3 \varepsilon \int \di^3 x \sqrt{-g} |\partial\chi| \partial_\mu\chi \partial^\mu \hat\Box\chi = 3 \varepsilon \int \di^3 x \partial_\nu (g^{\mu\nu}\sqrt{-g} |\partial\chi| \partial_\mu\chi) \hat\Box\chi \\ 
&&\quad= 3 \varepsilon \int \di^3 x \hat\partial_\nu (\hat g^{\mu\nu}\sqrt{-\hat g} \hat\partial_\mu\chi) \hat\Box\chi= 3 \varepsilon \int \di^3 x \sqrt{-\hat g} \hat \Box \chi \hat \Box \chi \,.
\eea
Therefore one can choose $\varepsilon$ to eliminate the additional operator.\footnote{We checked explicitly that this extra operator cancels when calculating correlation functions of $J^\mu$ and $T^{\mu\nu}$.}

Let us now move to next order, adding two derivatives more. We will disregard operators that start cubic or higher in perturbations $\pi$ and $\delta g_{\mu\nu}$: since our positivity bounds derive from two-point functions, operators that start cubic or higher will not play any role. First of all, let us consider what can be eliminated using field redefinitions analogous to Eq.~\eqref{eq:fieldred}, but with two more derivatives. In analogy with the previous order, once we plug the new field redefinition in the leading Lagrangian, Eq.~\eqref{eq:S1}, this will generate a term proportional to the leading-order equation of motion, $\hat\Box \chi$, multiplied by a scalar quantity built with the $\hat g_{\mu\nu}$ metric. This implies that one can dispose of all operators that include a $\hat \Box\chi$.

One simplification in $d=3$ is that the Riemann tensor can be written in terms of the Ricci tensor. Let us start with operators with two Ricci tensors (it is not possible to have more since there would be too many derivatives). These can be written as
\begin{equation}
\hat R^2\,, \quad \hat R_{\mu\nu} \hat R^{\mu\nu} \,, \quad \hat R_{\mu}^0 \hat R^{\mu0} \,, \quad \hat R \hat R^{00} \,, \quad \hat R^{00} \hat R^{00} \,. 
\end{equation}
We are using the following notation: if a tensor appears with an upper $0$ index, it means this index is contracted with $\hat\partial\chi$. For instance, in this notation the operator proportional to $c_3$ is the action~\eqref{eq:S2} would be written as $c_3 \hat R^{00}$. (In a unitary gauge in which $\pi =0$ one has $\hat\partial_\alpha\chi = \mu \, \delta_\alpha^0$, hence the notation.\footnote{This notation is reminiscent of the Effective Field Theory of Inflation \cite{Cheung:2007st}. In fact the Lagrangian we are studying here is a particular case of this, with additional symmetries.}) It turns out one can dispose of the last two operators. One can write
\begin{equation}
\hat R^{00} = \hat R^{\alpha\beta} \hat\partial_\alpha\chi\hat\partial_\beta\chi = \hat g_{\mu\nu} \hat R^{\mu\alpha\nu\beta} \hat\partial_\alpha\chi\hat\partial_\beta\chi = \hat g^{\mu\nu}  \left[ (\hat\nabla_\mu\hat\nabla_\alpha-\hat\nabla_\alpha \hat\nabla_\mu) \hat\partial_\nu\chi \right]    \hat\partial^\alpha\chi  \sim \hat g^{\mu\nu}  \left[ \hat\nabla_\mu\hat\nabla_\nu \hat\partial_\alpha\chi \right]    \hat\partial^\alpha\chi  \,,
\end{equation}
where in the last step we dropped a term that contains $\hat\Box\chi$ since, as we discussed, these terms can be removed by a field redefinition. We also exchanged the indices $\alpha$ and $\nu$ in the remaining term, since the derivatives are acting on a scalar. Now notice that if the last term $\hat \partial^\alpha \chi$ were inside the covariant derivatives, one would get $(\hat\partial\chi)^2 = 1$ and so the term would vanish. Therefore $\hat R^{00}$ does not vanish, but it can be written schematically as $(\hat\nabla\hat\nabla\chi)^2$. This object is already quadratic since $\hat\nabla\hat\nabla\chi$ vanishes on the background. Therefore $\hat R \hat R^{00} $ and $\hat R^{00} \hat R^{00} $ are at least cubic and we can disregard them.

Let us now consider terms with a single Ricci tensor. We can assume without loss of generality that there are no derivatives acting on $\hat R$, otherwise one integrates by parts. Terms of the schematic structure $\hat R \hat\nabla\hat\nabla\chi \hat\nabla\hat\nabla\chi$ start at cubic order, so we need to focus on the terms 
\begin{equation}\label{eq:R3d}
\hat R^{00} \hat \nabla^0 \hat \nabla^0 \hat \nabla^0 \chi \,, \quad  \hat R  \hat \nabla^0 \hat \nabla^0 \hat \nabla^0 \chi \,, \quad \hat R^{\mu\nu}  \hat \nabla_\mu \hat \nabla_\nu \hat \nabla^0 \chi  \,, \quad \hat R^{0\mu} \hat \nabla_\mu \hat \nabla^0 \hat \nabla^0 \chi \,.
\end{equation}
Notice that we already dropped terms that contain $\hat\Box\chi$ and that the order of the derivatives does not count, since in commuting them we would generate terms with two Ricci's, but these were studied above.  The last two terms can be disposed of, using the contracted Bianchi identity: $\nabla_\mu (R^{\mu}_{\nu}-\frac{1}{2} R g_{\mu\nu})=0$. For instance, integrating by parts,
\begin{equation}
\hat R^{0\mu} \hat \nabla_\mu \hat \nabla^0 \hat \nabla^0 \chi  \sim - \hat \nabla_\mu \hat R^{0\mu} \hat \nabla^0 \hat \nabla^0 \chi = \frac12 \hat \nabla^0 \hat R \hat \nabla^0 \hat \nabla^0 \chi \sim -\frac12 \hat R \hat \nabla^0  \hat \nabla^0 \hat \nabla^0 \chi \,,
\end{equation}
so that the fourth operator becomes proportional to the second. The same holds for the third one, which gives rise to a term that contains $\hat\Box\chi$. (Notice that in doing integration by parts, sometimes the derivative acts on the ``hidden'' $\hat\partial\chi$ which is implicit when we have upper index $0$ in our notation. In this case, however, one generates terms of the schematic form $\hat R \hat\nabla\hat\nabla\chi \hat\nabla\hat\nabla\chi$, and these start cubic as we discussed.)

We now want to show that the object $\hat \nabla^0 \hat \nabla^0 \hat \nabla^0 \chi$ is actually quadratic in perturbations and this will imply that the the first two terms in Eq.~\eqref{eq:R3d} are cubic and we can also discard them. Explicitly
\begin{equation}\label{eq:000}
\hat \nabla^0 \hat \nabla^0 \hat \nabla^0 \chi  =  \left( \hat \nabla^\alpha \hat \nabla^\beta \hat \nabla^\gamma \chi\right) \hat\partial_\alpha \chi \hat\partial_\beta \chi \hat\partial_\gamma \chi\,.
\end{equation} 
If $\hat\partial_\gamma\chi$ were inside the $\hat\nabla^\alpha \hat\nabla^\beta$, it would contract to give $\hat\partial^\gamma \chi\hat\partial_\gamma \chi= 1$, so that the whole term would vanish. The difference between this and the expression~\eqref{eq:000} involves derivatives acting on $\hat\partial_\gamma\chi$, but these extra terms are quadratic: we conclude that $\hat \nabla^0 \hat \nabla^0 \hat \nabla^0 \chi$ is quadratic. In conclusion we can dispose of all terms in Eq.~\eqref{eq:R3d}.

Finally, we have to consider terms without any Ricci tensor. Since we are not interested in terms which are cubic or higher, we only consider operators of the schematic form $(\hat\nabla\hat\nabla\hat\nabla\chi)^2$ (the others can be brought to this form integrating by parts). The order of derivatives is immaterial, since in commuting one would generate terms involving Ricci and these were already considered. If two indices are contracted, one can always commute and integrate by parts to generate a term that contains $\hat\Box\chi$ and this can be set to zero by a field redefinition. The other possibility is that none of the indices are contracted and they are all upper $0$'s. But we just proved $\hat \nabla^0 \hat \nabla^0 \hat \nabla^0 \chi$ is quadratic, so also this possibility gives an operator that is cubic or higher. 

In conclusion, the only operators with four additional derivatives that start quadratic in perturbations are
\begin{equation}
S^{(3)} =  \int \di^3 x \sqrt{-\hat g} \left(c_4  \hat R^2 + c_5 \hat R_{\mu\nu} \hat R^{\mu\nu}  + c_6 \hat R_{\mu}^0 \hat R^{\mu0}  \right) \,. 
\end{equation}

\section{Conservation laws for $J^\mu$ and $T^{\mu\nu}$\label{app:conservation}}
It is interesting to review if and how the conservation laws of currents are satisfied within correlation functions. We shall focus on the retarded Green's functions as defined through the functional derivative of the path integral in Eq.~\eqref{eq:GRfunctional2}. We will review that contact terms and the fact that currents take a vev make the discussion quite subtle.

Let us start with the $U$(1) current and consider the following path integral: 
\bea\label{eq:K1}
{\cal{K}}=\int {\cal{D}}\phi\; e^{i\,\int \di^d x\; {\cal{L}}\left(\phi,A_\mu\right)} \ .
\eea
Now, we can change variables to a gauge-transformed field $\phi'=e^{-i\alpha(x)}\phi$. Since the measure of integration is invariant (assuming a non-anomalous symmetry), ${\cal{D}}\phi'={\cal{D}}\phi$, we have
\bea
{\cal{K}}=\int {\cal{D}}\phi\; e^{i\,\int \di^d x\; {\cal{L}}\left(\phi'(\phi),A_\mu\right)}\ .
\eea
The fact that the Lagrangian is gauge invariant gives  ${\cal{L}}\left(\phi'(\phi),A_\mu-\dd_\mu\alpha\right)={\cal{L}}\left(\phi,A_\mu\right)$, which we can also write as ${\cal{L}}\left(\phi'(\phi),A_\mu\right)={\cal{L}}\left(\phi,A_\mu+\dd_\mu\alpha\right)$. So, assuming infinitesimal $\alpha$, we can write
\bea
{\cal{K}}=\int {\cal{D}}\phi\; e^{i\,\int \di^d x\; {\cal{L}}\left(\phi,A_\mu+\dd_\mu\alpha\right)}=\int {\cal{D}}\phi\; e^{i\,\int \di^d x\; {\cal{L}}\left(\phi,A_\mu\right)} \left(1+i \int \di^d x\; \dd_\mu\alpha(x) \frac{\delta{S}}{\delta A_\mu(x)}\right)\ .
\eea
Equating this to~\eqref{eq:K1}, we get, upon integration by parts,
\bea\nn
&& 0=\int {\cal{D}}\phi\; e^{i\,\int \di^d x\; {\cal{L}}\left(\phi,A_\mu\right)} \int \di^dx\;\dd_\mu \alpha(x) \frac{\delta{S}}{\delta A_\mu(x)}\\ \nn
&&\quad=-\int \di^dx\;\alpha(x) \;\dd_{x^\mu} \int {\cal{D}}\phi\; e^{i\,\int \di^d x'\; {\cal{L}}\left(\phi(x'),A_\nu(x')\right)}   \frac{\delta{S}}{\delta A_\mu(x)}\\ 
&&\quad=i\int \di^dx\;\alpha(x) \;\dd_{x^\mu}  \frac{\delta}{\delta A_\mu(x)} \int {\cal{D}}\phi\; e^{i\,\int \di^d x'\; {\cal{L}}\left(\phi(x'),A_\nu(x')\right)}   \ .
\eea
Since this must be true for every $\alpha(x)$, we obtain
\be\label{eq:K2}
0=\dd_{x^\mu}\frac{\delta}{{\delta A_\mu(x)}}\int {\cal{D}}\phi\; e^{i\,\int \di^d x'\; {\cal{L}}\left(\phi(x'),A_\nu(x')\right)} \ .
\ee
We can take a second derivative with respect to $A_\nu(y)$, and then set $A_\mu=0$, to get
\be\label{eq:K3}
0=\left.\dd_{x^\mu}\frac{\delta^2}{{\delta A_\mu(x)}{\delta A_\nu(y)}}\int {\cal{D}}\phi\; e^{i\,\int \di^d x'\; {\cal{L}}\left(\phi(x'),A_\rho(x')\right)}\right|_{A_\sigma=0} \ .
\ee
Eq.~\eqref{eq:K2} (evaluated at $A_\mu=0$) and~\eqref{eq:K3} are exactly the functional derivatives that enter in the computation of the Green's function of the $U$(1) currents $\langle J^\mu(-k) J^\nu(k)\rangle$, as derived in~\eqref{eq:GRfunctional2} (which is not to be confused with correlation functions of Noether currents, with respect to which it differs by contact terms). We therefore conclude that  $\langle J^\mu(-k) J^\nu(k)\rangle$ is exactly conserved, without the presence, on the RHS, of any $\delta$-function term. This justifies the tensorial structure that we assumed (and verified) in~\eqref{JmuJnu}. 

We now perform the same computation for the Green's function of the stress-energy tensor (again, meant as defined by~\eqref{eq:GRfunctional2}). The same manipulations lead now to
\be\label{eq:K4}
0=\int {\cal{D}}\phi\; e^{i\,\int \di^d x'\;\sqrt{-g}\; {\cal{L}}\left(\phi(x'),g_{\rho\sigma}(x')\right)}  \int \di^dx\;\sqrt{-g} \frac{1}{\sqrt{-g}}\frac{\delta S}{ \delta g_{\mu\nu}} \nabla_{(\mu}\xi_{\nu)}\ ,
\ee
where $\xi_\nu$ is the parameter of an infinitesimal diffeomorphism, and round brackets stand for symmetrization. This leads to the following identity:
\bea\label{eq:K5}
&&0=-i \nabla_{x^\mu}\int {\cal{D}}\phi\; e^{i\,\int \di^d x'\;\sqrt{-g}\; {\cal{L}}\left(\phi(x'),g_{\rho\sigma}(x')\right)} \left(\frac{1}{\sqrt{-g(x)}}\frac{\delta S}{ \delta g_{\mu\nu}(x)}\right)=\\ \nn
&&\quad=\nabla_{x^\mu} \left(\frac{1}{\sqrt{-g(x)}}\frac{\delta }{ \delta g_{\mu\nu}(x)} \int {\cal{D}}\phi\; e^{i\,\int \di^d x'\;\sqrt{-g}\; {\cal{L}}\left(\phi(x'),g_{\rho\sigma}(x')\right)}\right) \ .
\eea
Now, interestingly, if we act with a second derivative with respect to the metric, this second derivative will act also on the Christoffel symbols associated to the covariant derivative. Explicitly, using some Christoffel symbol's identities, we can write
\bea\label{eq:K6}
&&0=\frac{1}{\sqrt{-g(y)}} \frac{\delta}{\delta g_{\rho\sigma}(y)}\nabla_{x^\mu} \left(\frac{1}{\sqrt{-g(x)}}\frac{\delta }{ \delta g_{\mu\nu}(x)} \int {\cal{D}}\phi\; e^{i\,\int \di^d x'\;\sqrt{-g}\; {\cal{L}}\left(\phi(x'),g_{\alpha\beta}(x')\right)}\right) \\ \nn
&&=\nabla_{x^\mu} \left(\frac{1}{\sqrt{(-g(x)(-g(y)) }}\frac{\delta^2 }{ \delta g_{\mu\nu}(x)\delta g_{\rho\sigma}(y)} \int {\cal{D}}\phi\; e^{i\,\int \di^d x'\;\sqrt{-g}\; {\cal{L}}\left(\phi(x'),g_{\alpha\beta}(x')\right)}\right)\\  \nn
&&\quad+\frac{1}{\sqrt{-g(y)}} \frac{\delta}{\delta g_{\rho\sigma}(y)}\left( \frac{1}{\sqrt{-g(x)}} \Gamma^\nu_{\theta\gamma}(x)\right)\left( \frac{\delta}{\delta g_{\theta\gamma}} \int {\cal{D}}\phi\; e^{i\,\int \di^d x'\;\sqrt{-g}\; {\cal{L}}\left(\phi(x'),g_{\alpha\beta}(x')\right)} \right)\ .
\eea
Evaluating this expression at $g_{\mu\nu}=\eta_{\mu\nu}$, we obtain
\bea\label{eq:K7}
&&0=\dd_{x^\mu} \left.\left(\frac{\delta^2 }{ \delta g_{\mu\nu}(x)\delta g_{\rho\sigma}(y)} \int {\cal{D}}\phi\; e^{i\,\int \di^d x'\;\sqrt{-g}\; {\cal{L}}\left(\phi(x'),g_{\alpha\beta}(x')\right)}\right)\right|_{g_{\alpha\beta}=\eta_{\alpha\beta}}\\  \nn
&&\quad+ \frac{1}{\sqrt{-g(x)}} \frac{\delta \Gamma^\nu_{\theta\gamma}(x) }{\delta g_{\rho\sigma}(y)}\left.  \Big |_{g_{\mu\nu}=\eta_{\mu\nu}} \cdot \left(\frac{\delta}{\delta g_{\theta\gamma}} \int {\cal{D}}\phi\; e^{i\,\int \di^d x'\;\sqrt{-g}\; {\cal{L}}\left(\phi(x'),g_{\alpha\beta}(x')\right)}\right)\right|_{g_{\mu\nu}=\eta_{\mu\nu}} \ .
\eea
The term on the second line introduces some $\delta$-function terms ({\it i.e.} proportional to $\delta^{(d)}(x-y)$) in the conservation equation $\dd_\mu \langle T^{\mu\nu}T^{\rho\sigma}\rangle$. Notice that this term vanishes unless there is a vacuum expectation value for the stress tensor: $\langle T^{\theta\gamma}\rangle\neq 0$. This is the case for the theory we study in \S\ref{superfluidsec}, for a term proportional to $c_1$.  The tensorial structure assumed in~\eqref{Tmunugeneral} is therefore violated by a term in $c_1$ (which was irrelevant for the discussion there). We verified that the conservation law in~\eqref{eq:K7} is indeed satisfied by the full answer we obtained for $\langle T^{\mu\nu}T^{\rho\sigma}\rangle$.

\section{Contour argument in the upper half plane} \label{k0sec}
In this appendix we want to show that one can draw the same conclusions as in the main text with a contour argument that remains in the upper half of the complex plane. Indeed, instead of considering the two contours of Fig.~\ref{fig:doublecontour}, one can just concentrate on the upper one.

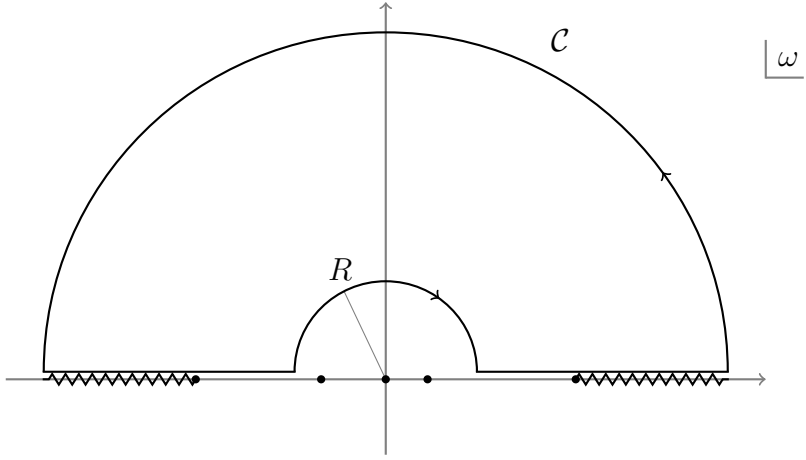
\begin{figure}[h!]
\centering
\begin{tikzpicture}[thick]

  \draw[->,gray] (-5,0) -- (5,0);
  \draw [->,gray] (0,-1) -- (0,5);

  \draw[gray, thick] (5,4) -- (5.5,4);
  \draw[gray, thick] (5,3.987) -- (5,4.5);
  \node at (5.3,4.25) {$\omega$};
        
        \node at (2.3,4.5) {$\mathcal{C}$};
        
        \draw[black,
        decoration={markings, mark=at position 0.2 with {\arrow{>}}},
        postaction={decorate}] (4.5,0.1) arc
    [
        start angle=0,
        end angle=180,
        x radius=4.5cm,
        y radius =4.5cm
    ];
    
    \draw[gray,thin] (0,0) -- (-0.55,1.17);
    \node at (-0.6,1.46) {$R$};
    
    \draw[black] (-4.514,0.1) -- (-1.2861+0.1,0.1);
    
    \draw[black,decoration={markings, mark=at position 0.7 with {\arrow{>}}},
        postaction={decorate}] (-1.2,0.1) arc
    [
        start angle=180,
        end angle=0,
        x radius=1.2cm,
        y radius=1.2cm
    ];
    
    \draw[black] (1.086+0.1,0.1) -- (4.514,0.1);
    \draw[black,fill=black] (0,0) circle (0.04cm);
    \draw[black,fill=black] (-1.05+0.2,0) circle (0.04cm);
    \draw[black,fill=black] (0.85-0.3,0) circle (0.04cm);
    
    \draw[decorate,decoration={zigzag, segment length=5, amplitude=2}] (-2.5,0) -- (-4.514,0);
    \draw[black,fill=black] (2.5,0,0) circle (0.04cm);
    \draw[decorate,decoration={zigzag, segment length=5, amplitude=2}] (2.5,0) -- (4.514,0);
    \draw[black,fill=black] (-2.5,0,0) circle (0.04cm);
 \end{tikzpicture}
\caption{The contour $\mathcal{C}$ that is used in Eq.~\eqref{Cauchy}.} \label{thecontour}
\end{figure}

Consider the function
\begin{equation} \label{ftilde}
	\tilde{f}(\omega) = \tilde{G}_R^{\mu \nu}(\omega,\boldsymbol{k}_0+\omega \boldsymbol{\xi}) V_\mu(\omega) V_\nu(\omega^*)^* \,,
\end{equation}
where $\boldsymbol{k}_0, \boldsymbol{\xi}$ are arbitrary real $(d-1)$-vectors with $\boldsymbol{\xi}^2 < 1$, and $V^\mu(\omega)$ are polynomials. Then $\tilde{f}$ is analytic in the upper half $\omega$-plane, $\omega^\textsf{Im} > 0$. For any integer $\ell$, Cauchy's theorem tells us that
\begin{equation} \label{Cauchy}
	\int_\mathcal{C} \frac{\di \omega}{\omega^\ell} \tilde{f}(\omega) = 0 \,,
\end{equation}
where $\mathcal{C}$ is any closed curve that lies completely in the upper half plane. We choose $\mathcal{C}$ as in Fig.~\ref{thecontour}, so that the integral on the LHS of~\eqref{Cauchy} can be expressed as a sum of contributions from the real line $+ i \varepsilon$, avoiding the segment $[-R,R]$ by passing along a semicircle of radius $R$, and a large ``arc at infinity''.

We assume $\ell \geq 1$ is large enough so that the contribution from the arc at infinity vanishes, and additionally that it is odd (this is necessary for positivity, see below).
Then we may express~\eqref{Cauchy} as
\begin{align} \label{AAA}
	0 = \int_R^\infty \frac{\di \omega}{\omega^\ell} \left( \tilde{f}(\omega + i \varepsilon) - \tilde{f}(-\omega + i \varepsilon) \right) + A_\textsf{tree}(R) + A_\textsf{loop}(R) \,.
\end{align}
Here we have done a change of variables $\omega \rightarrow -\omega$ for the integral along the negative real axis, using that $\ell$ is odd and we have, within the disc of radius $R$ centered at the origin, split up $\tilde{f}$ in tree-level and loop contributions,
\begin{equation} \label{fsplit}
	\tilde{f}(\omega) = \tilde{f}_\textsf{tree}(\omega) + \tilde{f}_\textsf{loop}(\omega) \,.
\end{equation}
$A_\textsf{tree,loop}$ are the contributions from $\tilde{f}_\textsf{tree,loop}$ respectively to the integral along the finite semicircle. We assume that $\tilde{f}_\textsf{tree}(\omega)$ is meromorphic in a region containing low frequencies $|\omega| < R$, with as its possible singularities only simple poles at real $\omega_i \neq 0$. Notice that $\tilde{f}_\textsf{tree}(\omega)$, away from the poles, is real for real $\omega \in \mathbb{R}$ as a consequence of Eq.~\eqref{starstar}. If the tree-level approximation of the EFT is good, then $A_\textsf{loop}(R)$ is a small correction to $A_\textsf{tree}(R)$.

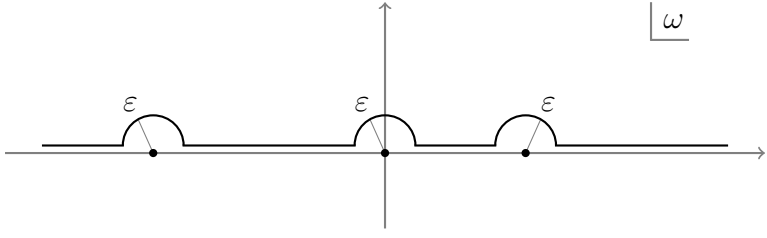
\begin{figure}[h!]
\centering
\begin{tikzpicture}[thick]

  \draw[->,gray] (-5,0) -- (5,0);
  \draw [->,gray] (0,-1) -- (0,2);

  \draw[gray, thick] (5-1.5,1.5) -- (5.5-1.5,1.5);
  \draw[gray, thick] (5-1.5,3.987-2.5) -- (5-1.5,4.5-2.5);
  \node at (5.3-1.5,4.25-2.5) {$\omega$};
    
    \draw[gray,thin] (0,0) -- (-0.2,0.45);
    \node at (-0.30,0.65) {$\varepsilon$};
    
    \draw[gray,thin] (-1.05-2,0) -- (-1.05-2-0.2,0.45);
    \node at (-1.05-2-0.3,0.65) {$\varepsilon$};
    
    \draw[gray,thin] (0.85+1,0) -- (0.85+1.2,0.45);
    \node at (0.85+1.3,0.65) {$\varepsilon$};
    
    \draw[black] (-4.514,0.1) -- (-3.436,0.1);
    
    \draw[black] (-0.4,0.1) arc
    [
        start angle=180,
        end angle=0,
        x radius=0.4cm,
        y radius=0.4cm
    ];
    
    \draw[black] (-1.05-2-0.4,0.1) arc
    [
        start angle=180,
        end angle=0,
        x radius=0.4cm,
        y radius=0.4cm
    ];
    
    \draw[black] (0.85+1-0.4,0.1) arc
    [
        start angle=180,
        end angle=0,
        x radius=0.4cm,
        y radius=0.4cm
    ];
    
    \draw[black] (0.386,0.1) -- (1.464,0.1);
    
    \draw[black] (-2.664,0.1) -- (-0.386,0.1);
    
    \draw[black] (1.086+1.15,0.1) -- (4.514,0.1);
    \draw[black,fill=black] (0,0) circle (0.04cm);
    \draw[black,fill=black] (-1.05-2,0) circle (0.04cm);
    \draw[black,fill=black] (0.85+1,0) circle (0.04cm);
 \end{tikzpicture}
\caption{The contour used to calculate $A_\textsf{tree}(R)$ in~\eqref{AAA}.} \label{treelevel}
\end{figure}

 We may calculate $A_\textsf{tree}(R)$ using the deformation in Fig.~\ref{treelevel}, giving

\begin{align}
	&A_\textsf{tree}(R) = \text{PV} \underset{[-R,R]}{\int} \frac{\di \omega}{\omega^\ell} \tilde{f}_\textsf{tree}(\omega) ~ - i \pi \sum_{\underset{\omega_i \neq 0}{\text{simple poles}}} \frac{\text{Res}(\tilde{f}_\textsf{tree},\omega_i)}{\omega_i^\ell} \\
	&-i \int_0^\pi \di t \frac{1}{\varepsilon^{\ell-1} e^{i(\ell-1)t}} \left( \tilde{f}_\textsf{tree}(0) + \tilde{f}_\textsf{tree}'(0) \varepsilon e^{it} + \frac{\tilde{f}_\textsf{tree}''(0)}{2!} \varepsilon^2 e^{2it} + \cdots + \frac{\tilde{f}_\textsf{tree}^{(\ell-1)}(0)}{(\ell-1)!} \varepsilon^{\ell-1} e^{i(\ell-1)t} + \mathcal{O}(\varepsilon^\ell) \right) \,, \notag
\end{align}
where the $\text{PV}$ is defined to avoid all the low-energy singularities of $\tilde{f}_\textsf{tree}/\omega^\ell$, as in Fig.~\ref{treelevel}. We have explicitly written out the integral of $\tilde{f}_\textsf{tree}/\omega^\ell$ along the small semicircle around $\omega = 0$, expanding $\tilde{f}_\textsf{tree}$ around this point. Since $\ell$ is odd, all the even-derivative terms vanish after integration, except the $(\ell-1)$st. This is independent of the value of $\boldsymbol{k}_0$ and the choice of $V^\mu(\omega)$. Notice that taking $\boldsymbol{k}_0 \neq \boldsymbol{0}$ and arbitrary polynomials $V^\mu(\omega)$ breaks the parity of $\tilde{f}_\textsf{tree}$ around $\omega = 0$, so that the odd-derivative terms do not vanish in general. We obtain
\begin{align}
	&A_\textsf{tree}(R) =\notag \\ &- i \pi \left( \frac{\tilde{f}_\textsf{tree}^{(\ell-1)}(0)}{(\ell-1)!} + \sum_{\underset{\omega_i \neq 0}{\text{simple poles}}} \frac{\text{Res}(\tilde{f}_\textsf{tree},\omega_i)}{\omega_i^\ell} \right) + \text{PV} \underset{[-R,R]}{\int} \frac{\di \omega}{\omega^\ell} \tilde{f}_\textsf{tree}(\omega) + 2 \sum_{\text{odd } k < \ell} \frac{\tilde{f}_\textsf{tree}^{(k)}(0)}{k!} \frac{\varepsilon^{k-\ell+1}}{k-\ell+1} + \mathcal{O}(\varepsilon) \,.
\end{align}
We will be interested in the imaginary part of this expression, which, since $\tilde{f}_\textsf{tree}$ is real on the real axis, is simply
\begin{equation}
	-\frac{1}{\pi} \, \text{Im} \, A_\textsf{tree}(R) = \frac{\tilde{f}_\textsf{tree}^{(\ell-1)}(0)}{(\ell-1)!} + \sum_{\underset{\omega_i \neq 0}{\text{simple poles}}} \frac{\text{Res}(\tilde{f}_\textsf{tree},\omega_i)}{\omega_i^\ell} \,.
\end{equation}

Now we turn to the principal value integral, which involves the full function $\tilde{f}$. Eq.~\eqref{GRAdiff} implies
\be
\tilde f (\omega +i \varepsilon) - \tilde f (\omega -i \varepsilon)  = i \times {\rm (positive)} \quad {\rm for} \quad\omega > 0 \ ,
\ee
while it is $i \;\times$ (negative) for $\omega<0$. Combining this with Eq.~\eqref{starstar} we get
\be
\begin{split}
\text{Im} \,\tilde f(\omega+ i \varepsilon) \geq 0  \quad {\rm for} \quad\omega > 0 \ ,\\
\text{Im} \,\tilde f(\omega + i \varepsilon) \leq 0  \quad {\rm for} \quad\omega < 0 \ .
\end{split}
\ee
This implies
\begin{align} \label{PVint}
	\text{Im} \int_R^\infty \frac{\di \omega}{\omega^\ell} &\left( \tilde{f}(\omega + i \varepsilon) - \tilde{f}(-\omega + i \varepsilon) \right)  \ge 0 \,.
\end{align}
Taking the imaginary part of~\eqref{AAA} and putting everything together, we obtain, schematically
\begin{align} \label{positivityeq}
	\frac{\tilde{f}_\textsf{tree}^{(\ell-1)}(0)}{(\ell-1)!} + \sum_{\underset{\omega_i \neq 0}{\text{simple poles}}} \frac{\text{Res}(\tilde{f}_\textsf{tree},\omega_i)}{\omega_i^\ell} - \frac{1}{\pi} \text{Im} \, A_\textsf{loop}(R) = (\text{positive}) \,.
\end{align}
We get exactly the same conclusions one would draw using the double contour of Fig.~\ref{fig:doublecontour} (it is easy to check that also the positive UV contributions match).

\section{Generalized Kramers-Kronig relation} \label{RKKsec}
As a classic example, consider the time-dependent electric susceptibility of a material $\chi_e$; it describes the (linear) relation between the electric field and the dielectric polarization density:
\be
\boldsymbol{P}(t) = \int_{-\infty}^{+\infty}\! \di t'\, \chi_e(t-t') \boldsymbol{E}(t')
\ee
or in Fourier space
\be
\boldsymbol{P}(\omega) = \tilde\chi_e(\omega) \boldsymbol{E}(\omega) \;.
\ee
Since the response of the material can only happen {\em after} the electric field is turned on, $\chi_e(t-t')$ vanishes for $t<t'$; it is retarded. This implies that $\tilde\chi_e(\omega)$ is analytic for $\omega$ in the upper half complex plane.
This property implies the classic Kramers-Kronig relation:
\begin{equation}
	\tilde{\chi}_e(\omega) = \frac{1}{i \pi} ~ \text{PV} \int_\mathbb{R} \frac{\di \zeta}{\zeta - \omega} \, \tilde{\chi_e}(\zeta) \,.
\end{equation}
Taking the real or imaginary parts of this relation, one can view the Kramers-Kronig relation as giving the real part of $\tilde\chi_e(\omega)$ in terms of its imaginary part and vice versa (see for instance \cite{Jackson:1998nia}). 

In the derivation above one assumed that the response of the material is non-local in time but local in space\footnote{The same behaviour occurs in the EFT of large scale structures, see for instance \cite{Carrasco:2012cv}.}, which is a good approximation for a non-relativistic medium. In general the response depends on space and time:  in this case one can generalize the Kramers-Kronig relation for functions that are not only retarded but also vanish outside the forward light cone \cite{MS76},
\begin{equation} \label{KKgen}
	\tilde{f}(\omega,\boldsymbol{k}) = \frac{1}{i \pi} \, \text{PV} \int_\mathbb{R} \frac{\di \zeta}{\zeta - \omega} \tilde{f}\left( \zeta,\boldsymbol{k} + (\zeta-\omega)\boldsymbol{\xi} \right) \,.
\end{equation}
(The ``ordinary'' Kramers-Kronig relation is the one with $\boldsymbol{\xi} = \boldsymbol{0}$.) Here $\tilde{f}(k)$ is the Fourier transform of a response function $f(x)$ which vanishes outside the forward lightcone in $d$ dimensions, and $\boldsymbol{\xi}$ is any real $(d-1)$-vector with $\boldsymbol{\xi}^2 < 1$. Setting $\boldsymbol{k} = \boldsymbol{k}_0 + \omega \boldsymbol{\xi}$ for some $\boldsymbol{k}_0 \in \R^{d-1}$ we obtain a more symmetric-looking relation
\begin{equation} \label{RKKbis}
	\tilde{f}(\omega,\boldsymbol{k}_0 + \omega \boldsymbol{\xi}) = \frac{1}{i \pi} \, \text{PV} \int_\mathbb{R} \frac{\di \zeta}{\zeta - \omega} \tilde{f}\left( \zeta,\boldsymbol{k}_0 + \zeta\boldsymbol{\xi} \right) \,.
\end{equation}

In this appendix we will show that the relation~\eqref{RKKbis} has, at tree level, the same positivity implication as the contour integral argument we presented in Appendix \ref{k0sec}. So we will assume here that at low energies $|\omega| \leq R$, $\tilde{f}(\omega)$ is well-approximated by $\tilde{f}_\textsf{tree}(\omega)$ (cf.~\eqref{fsplit} and the assumptions below that equation), and correspondingly the term involving $A_\textsf{loop}(R)$ in~\eqref{positivityeq} will be absent in this section.

\subsubsection*{Derivation of~\eqref{KKgen}}
First, for convenience of the reader, we reiterate the derivation of~\eqref{KKgen} as it appears in \cite{MS76}. The vanishing of $f(x)$ outside the forward lightcone can be stated as
\begin{equation} \label{SR-RKK}
	f(t,\boldsymbol{x}) = \theta(t-\boldsymbol{\xi}\cdot\boldsymbol{x}) f(t,\boldsymbol{x})
\end{equation}
for all $\boldsymbol{\xi}$ with $\boldsymbol{\xi}^2 < 1$. (The surfaces $t = \boldsymbol{\xi}\cdot\boldsymbol{x}$ are codimension-one hyperplanes that lie outside of the forward and backward lightcones, intersecting them only at the origin. The collection of all these hyperplanes with $|\boldsymbol{\xi}| < 1$ are the spacelike-separated points from the origin.) The $d$-dimensional Fourier transform of $\theta(t-\boldsymbol{\xi}\cdot\boldsymbol{x})$ is
\begin{equation}
	\tilde{\theta}(\omega,\boldsymbol{k};\boldsymbol{\xi}) = \frac{i  (2\pi)^{d-1} \, \delta^{(d-1)}(\boldsymbol{k} - \omega \boldsymbol{\xi})}{\omega + i \varepsilon}
\end{equation}
with $\varepsilon \rightarrow 0^+$, because indeed
\begin{equation}
	(2 \pi)^{-d} \int_\mathbb{R} \di \omega \int_{\mathbb{R}^{d-1}} \di^{d-1} \boldsymbol{k} \, e^{-i \omega t + i \boldsymbol{k}\cdot \boldsymbol{x}} \tilde{\theta}(\omega,\boldsymbol{k};\boldsymbol{\xi}) = \frac{i}{2 \pi} \int_\mathbb{R} \di \omega \, \frac{e^{-i \omega(t - \boldsymbol{\xi}\cdot\boldsymbol{x})}}{\omega + i \varepsilon} = \theta(t-\boldsymbol{\xi}\cdot\boldsymbol{x}) \,.
\end{equation}
Using the convolution theorem on the product~\eqref{SR-RKK}, and the Sokhotski-Plemelj identity
\begin{equation} \label{sokhotskiplemelj}
	\frac{1}{\omega + i \varepsilon} = -i \pi \, \delta(\omega) + \text{PV}\left( \frac{1}{\omega} \right) \,,
\end{equation}
one readily arrives at~\eqref{KKgen}:
\begin{align}\label{eq:KKderived}
	\tilde{f}(\omega,\boldsymbol{k}) &= \int_{\R^d} \frac{\di^d k'}{(2\pi)^d} \frac{(2\pi)^{d-1} i}{\omega - \omega' + i \varepsilon} \delta^{(d-1)}\left( \boldsymbol{k} - \boldsymbol{k}' - (\omega-\omega')\boldsymbol{\xi} \right) \tilde{f}(\omega',\boldsymbol{k}') \notag \\
	&= \frac{i}{2\pi} \int_\R \di \omega' \frac{1}{\omega - \omega' + i \varepsilon} \, \tilde{f} \left( \omega',\boldsymbol{k}+(\omega'-\omega)\boldsymbol{\xi} \right) \notag \\
	&= \frac{1}{2} \tilde{f}(\omega,\boldsymbol{k}) + \frac{1}{2\pi i} \text{PV} \int_\R \frac{\di \omega'}{\omega' - \omega} \, \tilde{f} \left( \omega',\boldsymbol{k}+(\omega'-\omega)\boldsymbol{\xi} \right) \,,
\end{align}
from which the result follows.

An alternative proof of~\eqref{RKKbis} starts by noticing that the function $\tilde{f}(\omega,\boldsymbol{k}_0 + \omega \boldsymbol{\xi})$ is analytic as a function of $\omega$ in the upper half complex plane, as a consequence of the $\theta$-function in Eq.~\eqref{SR-RKK}. Then one can follow the usual textbook derivation of the Kramers-Kronig relation. One considers $\tilde{f}(\zeta,\boldsymbol{k}_0 + \zeta \boldsymbol{\xi})/(\zeta-\omega + i \varepsilon)$: this is analytic in the upper $\zeta$-plane. Therefore one gets zero integrating all along the whole real $\zeta$-axis and closing the contour with an arc at infinity in the upper $\zeta$-plane. If the function $\tilde f$ decays at infinity sufficiently fast, the contribution of the arc vanishes and Eq.~\eqref{RKKbis} follows using Eq.~\eqref{sokhotskiplemelj}.

\subsubsection*{Illustration of~\eqref{RKKbis}}
We can illustrate~\eqref{RKKbis} in the simplest example of a Lorentz-invariant scalar of mass $m$. This will allow us to illustrate a subtlety in the application of the Kramers-Kronig relations. The retarded Green's function (indeed a function that vanishes outside the forward light cone) in Fourier space reads
\begin{equation}\label{GRrel}
	\tilde{G}_R(\omega,\boldsymbol{k}) = \frac{1}{(\omega+i\varepsilon)^2 - \boldsymbol{k}^2 - m^2} = \text{PV} \left( \frac{1}{\omega^2 - \boldsymbol{k}^2 - m^2} \right) + \frac{i \pi}{2 E_{\boldsymbol{k}}} \left( \delta \left( \omega + E_{\boldsymbol{k}} \right) - \delta \left( \omega - E_{\boldsymbol{k}} \right) \right) \,,
\end{equation}
where $E_{\boldsymbol{k}} \equiv \sqrt{\boldsymbol{k}^2 + m^2}$, and, similarly,
\begin{equation}
	\tilde{G}_R(\zeta,\boldsymbol{k} + \zeta\boldsymbol{\xi}) = \text{PV} \left( \frac{1}{\zeta^2 - (\boldsymbol{k} + \zeta\boldsymbol{\xi})^2 - m^2} \right) + \frac{i \pi}{(1-\boldsymbol{\xi}^2) (\zeta_+ - \zeta_-)} \left( \delta(\zeta-\zeta_-) - \delta(\zeta - \zeta_+) \right) \,,
\end{equation}
where we decomposed $\zeta^2 - (\boldsymbol{k} + \zeta\boldsymbol{\xi})^2 - m^2 = (1-\boldsymbol{\xi}^2)(\zeta-\zeta_+)(\zeta-\zeta_-)$, with
\begin{align}
	\zeta_\pm = \frac{\boldsymbol{\xi} \cdot \boldsymbol{k} \pm \sqrt{(1-\boldsymbol{\xi}^2)\left(\boldsymbol{k}^2+m^2\right) + (\boldsymbol{\xi} \cdot \boldsymbol{k})^2}}{1-\boldsymbol{\xi}^2} \,.
\end{align}
(Notice $\zeta_- < 0 < \zeta_+$.) To verify~\eqref{RKKbis}, the contribution to the RHS from the imaginary part of $\tilde{G}_R$ is
\be
\frac{1}{(1-\boldsymbol{\xi}^2) (\zeta_+ - \zeta_-)} \left( \frac{1}{\zeta_- - \omega} - \frac{1}{\zeta_+ - \omega} \right)  =  \frac{1}{(1-\boldsymbol{\xi}^2) \left( \omega - \zeta_+ \right) \left( \omega - \zeta_- \right)}= \frac{1}{\omega^2 - (\boldsymbol{k}+ \omega \boldsymbol{\xi})^2 - m^2} \,.
\ee
Notice that the principal-value prescription in the $\zeta$-integral of Eq.~\eqref{RKKbis} instructs that also this final expression must be interpreted in principal value for $\omega$ (this is clear if, before integrating in $\zeta$, we consider to integrate in $\omega$ against a smooth test function). Therefore we get indeed the real part of the LHS.

It remains to be shown that
\begin{align} \label{lastthing}
\frac{\pi}{(1-\boldsymbol{\xi}^2) (\zeta_+ - \zeta_-)} \left( \delta(\omega-\zeta_-) - \delta(\omega - \zeta_+) \right) = \frac{-1}{\pi \left( 1 - \boldsymbol{\xi}^2 \right)} \text{PV} \int_{\mathbb{R}} \frac{\di \zeta}{(\zeta-\omega)(\zeta-\zeta_+)(\zeta-\zeta_-)} \,.
\end{align}
A straightforward calculation shows that as long as $\omega \neq \zeta_\pm$, the integral on the RHS of~\eqref{lastthing} vanishes. The subtlety is what happens for $\omega = \zeta_{\pm}$. The point is that both in the definition of the Green's function, Eq.~\eqref{GRrel}, and in the derivation of the Kramers-Kronig relation, Eq.~\eqref{eq:KKderived}, one has to prescribe how to deal with the poles. When the two poles coincide, the prescriptions are ambiguous: one has to specify the order of the limits. In the derivation of the Kramers-Kronig identity one assume the analyticity of the Green's function in the upper half complex plane, so one is implicitly assuming a hierarchy between the two $\varepsilon$'s: with obvious notation $\varepsilon_{\rm Green} \gg \varepsilon_{\rm KK} $. Let us see that with this assumption we get the correct result. Let us concentrate on the case in which $\omega$ and $\zeta_+$ coincide (the case of $\zeta_-$ is analogous). Let us regulate the two simple poles in Eq.~\eqref{lastthing} as
\be
\int_{\mathbb{R}} \frac{\di \zeta}{(\zeta-\zeta_-)} \frac{(\zeta-\omega)}{(\zeta-\omega)^2 +\varepsilon_{\rm KK}^2} \frac{(\zeta-\zeta_+)}{(\zeta-\zeta_+)^2 +\varepsilon_{\rm Green}^2} \;.
\ee  
One can do the $\zeta$ integral and for infinitesimal $\varepsilon$'s with the hierarchy $\varepsilon_{\rm Green} \gg \varepsilon_{\rm KK}$ one gets
\be
 \frac{1}{(\zeta_+ -\zeta_-)} \frac{\pi \varepsilon_{\rm Green}}{(\omega-\zeta_+)^2+ \varepsilon_{\rm Green}^2} \sim \frac{1}{(\zeta_+ -\zeta_-)} \pi^2 \delta(\omega - \zeta_+) \;.
\ee
Plugging this back in Eq.~\eqref{lastthing} and considering the analogous $\zeta_-$ term one verifies the relation.

\subsubsection*{Equivalence of~\eqref{RKKbis} to contour integral argument at tree level}
The function $\tilde{f}$ in Eq.~\eqref{ftilde} is analytic for ${\rm Im}\;\omega > 0$, therefore it satisfies
\begin{equation} \label{RKKf}
	\tilde{f}(\omega) = \frac{1}{i \pi} \text{PV} \int_\mathbb{R} \frac{\di \zeta}{\zeta - \omega} \tilde{f}(\zeta) \,.
\end{equation}
For small $|\omega| \leq R$, we assume $\tilde{f}(\omega) \approx \tilde{f}_\textsf{tree}(\omega)$. In this regime we can take $\ell-1$ derivatives of both sides in~\eqref{RKKf} and evaluate the result at $\omega = 0$. Notice that since $\tilde f$ is real on the real axis at the regular points, the integral gives an imaginary contribution only at the poles of the function $\tilde f$ and on the cut of $\tilde f$. Assuming that $\ell$ is odd and equating the real parts, we obtain\footnote{Note that the integral in~\eqref{RKKf} could be divergent. We assume that formally taking enough derivatives renders it convergent.}
\begin{align}
	\frac{\tilde{f}_\textsf{tree}^{(\ell-1)}(0)}{(\ell-1)!} + \sum_{\underset{\omega_i \neq 0}{\text{simple poles}}} \frac{\text{Res}(\tilde{f}_\textsf{tree},\omega_i)}{\omega_i^\ell} \approx \frac{1}{\pi} \, \text{Im} \int_R^\infty \frac{\di \zeta}{\zeta^\ell} \left( \tilde{f}(\zeta+i \varepsilon) - \tilde{f}(-\zeta+i \varepsilon) \right) \,.
\end{align}
This is the same as Eq.~\eqref{positivityeq} in the tree-level approximation.

\section{UV theory with two scalars} \label{UV2scalars}
In this appendix we consider instead of the single-field UV theory~\eqref{UVLagrangian}, the two-field theory
\begin{align}
	\mathcal{L}_\textsf{UV} &= \sqrt{-g} \left( -|\partial \phi|^2 - (\partial \varphi)^2 - \lambda_1 |\phi|^6 - \lambda_2 \varphi^6 - \beta_1 |\phi|^2 \varphi^4 - \beta_2 |\phi|^4 \varphi^2 - \frac{R}{8} \left( |\phi|^2 + \varphi^2 \right) \right) \notag  \\
	&= \sqrt{-g} \left( -(\partial \rho)^2 + |\partial \theta|^2 \rho^2 - (\partial \varphi)^2 - \lambda_1 \rho^6 - \lambda_2 \varphi^6 - \beta_1 \rho^2 \varphi^4 - \beta_2 \rho^4 \varphi^2 - \frac{R}{8} \left( \rho^2 + \varphi^2 \right) \right) \,,
\end{align}
where $\phi = \rho \, e^{i \theta}$ is a complex scalar and $\varphi$ is a real one. As~\eqref{UVLagrangian} this theory is conformally invariant at tree level for all $\lambda_{1,2}$ and $\beta_{1,2}$. (It is necessary to have $\lambda_{1,2} \geq 0$ for the potential energy to be bounded below. There are further restrictions on $\beta_{1,2}, \lambda_{1,2}$ to be able to integrate out $\rho$ and $\varphi$ which will be discussed below.) The equations of motion for $\rho,\varphi$ are
\begin{align}
	\square \rho + |\partial \theta|^2 \rho - 3 \lambda_1 \rho^5 - \beta_1 \varphi^4 \rho - 2 \beta_2 \varphi^2 \rho^3 - \frac{R}{8} \rho &= 0 \,, \\
	\square \varphi - 3 \lambda_2 \varphi^5 - 2 \beta_1 \rho^2 \varphi^3 - \beta_2 \rho^4 \varphi - \frac{R}{8} \varphi &= 0 \,.
\end{align}
To integrate out $\rho$ and $\varphi$ we write the solution as $\rho = \rho_0 + \rho_1 + \cdots, \varphi = \varphi_0 + \varphi_1 + \cdots$ in a derivative expansion. As in \S\ref{3DUVsec} it suffices to know $\rho_{0,1}, \varphi_{0,1}$ to obtain the on-shell action to second subleading order in derivatives.

We assume initially that $\beta_1 \geq 0, \beta_2 \leq 0$. Then
\begin{align}
	\varphi_0^2 &= \frac{\beta_1}{3 \lambda_2} \left( \sqrt{1 - \frac{3 \lambda_2 \beta_2}{\beta_1^2}} - 1 \right) \rho_0^2 \,, \label{phi02} \\
	\rho_0^2 &= \frac{3\lambda_2}{\sqrt{2\beta_1^3 \left(1 - \sqrt{1 - 3 \beta_2 \lambda_2/\beta_1^2} \right) - 3 \beta_1 \beta_2 \lambda_2 \left( 3 - 2 \sqrt{1 - 3 \beta_2 \lambda_2/\beta_1^2} \right) + 27 \lambda_1 \lambda_2^2}} |\partial \theta| \,. \label{rho02}
\end{align}
We only have real solutions $(\rho_0,\varphi_0)$ when the quantity in the square root in the denominator of our expression for $\rho_0^2$ is positive.\footnote{This is not the case for all $\lambda_{1,2} > 0, \beta_1 \geq 0, \beta_2 \leq 0$, as one can easily check; take $\beta_1, \lambda_1 \ll 1$ and $|\beta_2|, \lambda_2 = \mathcal{O}(1)$ for instance. Also, notice one cannot expand around $(\rho_0,\varphi_0) = (0,0)$ because it is not a local minimum of the potential for any values of the parameters.} From here we may obtain expressions for $\rho_1, \varphi_1$ and the on-shell action, to second subleading order in derivatives. We leave out the details since the expressions are not particularly insightful. As in~\eqref{UVcoefficients} we find $c_ 3 = c_5 = c_6 = b = d = 0$, but the expressions for $c_{1,2,4}$ are altered. In contrast to the single-field theory, here the general EFT inequality~\eqref{4c4+2c5}, $c_4 \geq c_1^2/c_2$ when $c_3 = c_5 = c_6 = 0$, is not saturated:
\begin{equation}
	c_4 - \frac{c_1^2}{c_2} = \frac{\sqrt{2\beta_1^3 \left(1 - \sqrt{1 - 3 \beta_2 \lambda_2/\beta_1^2} \right) - 3 \beta_1 \beta_2 \lambda_2 \left( 3 - 2 \sqrt{1 - 3 \beta_2 \lambda_2/\beta_1^2} \right) + 27 \lambda_1 \lambda_2^2}}{768 \beta_1 \lambda_2 \sqrt{1 - 3 \beta_2 \lambda_2/\beta_1^2}} \,,
\end{equation}
which is generally strictly positive.

Now we turn to the case $\beta_1 \leq 0, \beta_2 \geq 0$. To have real solutions $(\rho_0,\varphi_0)$ we require $\beta_1^2 \geq 3 \lambda_2 \beta_2$. One of the solutions to the coupled equations for $\rho_0,\varphi_0$ in this case is also~\eqref{phi02}-\eqref{rho02}, but it is not a local minimum of the potential. Instead we must change the sign of the square roots in~\eqref{phi02}-\eqref{rho02}, resulting in a similar conclusion as before. The case $\beta_1 \leq 0, \beta_2 \leq 0$ is again similar. Finally when $\beta_1 \geq 0, \beta_2 \geq 0$ there are no solutions to expand about.

\newpage
\vfill\footnotesize
\bibliographystyle{klebphys2}
\bibliography{refs}
\end{document}